\documentstyle[seceq,preprint,mbf]{ptptex}

\def\slash#1{\ooalign{\hfil/\hfil\crcr$#1$}}

\def\XZ{{\mbf Z}}

\def\XQ{{\mbf Q}}
\def\XL{L}

\def\etI{A}
\def\etJ{B}
\def\etK{C}
\def\XM{{\mbf M}}

\def\XD{{\mbf D}}
\def\XS{{\mbf S}}

\def\XU{{\mbf U}}
\def\XT{T}
\def\XV{V}

\def\XG{{\mbf G}}

\def\dt{\!\cdot\!}
\def\nn{\nonumber\\}

\def\calL{{\cal L}}

\def\calD{{\cal D}}
\def\calN{{\cal N}}
\def\hatD{{\hat\calD}}

\def\slashD{{\hat{\slash\calD}}}

\def\hatR{{\hat R}}

\def\calP{{\cal P}}
\def\calR{{\cal R}}
\def\calT{{\cal T}}
\def\half{\hbox{\large ${1\over2}$}}
\def\myfrac#1#2{\hbox{\large ${#1\over#2}$}}

\def\pr#1{{#1'}}

\def\6#1{{\underline{#1}}}
\def\m6#1{{\underline{#1}\,}}
\def\Span#1{\setbox0=\hbox{$#1$}\rule[-\dp0]{0pt}{\ht0}}
\def\AB{{A}}
\def\BF{{\hat B}}
\def\cBF{{B}}
\def\GT{{A}}
\def\FA{{\cal H}}
\notypesetlogo  
\preprintnumber[3cm]{
KUNS-1801\\ 
hep-th/0208082}

\title{
Gauge and Non-Gauge Tensor Multiplets in 5D Conformal Supergravity
}

\author{
Taichiro {\sc Kugo}\footnote{E-mail:
kugo@gauge.scphys.kyoto-u.ac.jp}
and Keisuke {\sc Ohashi}\footnote{E-mail:
keisuke@gauge.scphys.kyoto-u.ac.jp}
}

\inst{
Department of Physics, Kyoto University, Kyoto 606-8502, Japan
}

\recdate{
\today
}

\abst{%
An off-shell formulation of two distinct tensor multiplets, a massive tensor 
multiplet and a tensor gauge multiplet, is presented
in superconformal tensor calculus in five-dimensional space-time. 
Both contain a rank 2 antisymmetric tensor field, but there is no gauge 
symmetry in the former, while it is a gauge field in the latter. Both 
multiplets have 4 bosonic and 4 fermionic {\em on-shell} modes, 
but the former consists of 16(boson)+16(fermion) component fields, while 
the latter consists of 8(boson)+8(fermion) component fields. }

\begin{document}
\maketitle

\section{Introduction}

Long ago, Townsend, Pilch and van Nieuwenhuizen\cite{ref:PTvN} 
discussed supersymmetry 
multiplets containing rank $2k$ massive antisymmetric tensor fields 
$\AB_{\mu_1\cdots\mu_{2k}}$ in odd numbers of dimensions $d=4k+1$ (and rank $2k-1$ fields 
in dimensions $d=4k-1$), which satisfy a self-dual type equations of 
motion: 
\begin{equation}
m\AB^{\mu_1\cdots\mu_{2k}}={i\over(2k)!}\epsilon^{\mu_1\cdots\mu_{2k}\nu_1\nu_2\cdots\nu_{2k+1}}
\partial_{\nu_1}\AB_{\nu_2\cdots\nu_{2k+1}}. \ 
\label{eq:selfdual}
\end{equation}
With the recent renewed interest in supergravity in five 
dimensions,\cite{ref:5Donshell} 
in connection with AdS/CFT 
dualities\cite{ref:AdS/CFT} 
and brane world scenarios,\cite{ref:HW,ref:hier,ref:RS} 
G\"unaydin and Zagermann\cite{ref:GZ} 
introduced tensor multiplets in 5D gauged supergravity system, 
generalizing the earlier work by G\"unayden, Sierra and 
Townsend\cite{ref:GST} on 5D supergravity system coupled to vector 
multiplets. They found that the presence of a tensor multiplet gives 
a novel contribution to the scalar potential. 
Very recently, Bergshoeff et~al.\cite{ref:BCWGHVP}\ 
have given a yet more general form of tensor-vector multiplet couplings 
in a 5D superconformal framework. 

These works all give the so-called {\em on-shell} formulation for the 
{\em massive} tensor multiplets, in which auxiliary fields are missing 
and the supersymmetry algebra closes only on-shell in a particular 
system. 

In a series of papers,\cite{ref:KO1,ref:KO2,ref:FO,ref:FKO,ref:KO3} 
 we have given a general 
off-shell formulation for 5D supergravity --- superconformal tensor 
calculus --- and have discussed the general supergravity system coupled 
to Yang-Mills vector multiplets and hypermultiplets in the 
5D bulk,\cite{ref:KO2} as 
well as its orbifold compactification on $S^1/Z_2$.\cite{ref:KO3} 

The work presented in
Ref.\citen{ref:KO1,ref:KO2,ref:FO,ref:FKO,ref:KO3} is, however,
incomplete, because the tensor multiplets are
missing, and therefore, in particular, 
the scalar potential is not quite general. 
This is because until now there had been no off-shell formulation of
the tensor multiplet. The purpose of this paper is to 
present such an off-shell formulation.
We have actually found an off-shell formulation not only of 
{\em massive (non-gauge)} tensor multiplets but also of massless
tensor {\em gauge} multiplets, both containing rank 2 tensor fields.
Although both multiplets have 4 bosonic and 4 fermionic 
{\em on-shell} modes, the former 
 consists of 16 (boson)+16 (fermion) component fields, while the 
latter consists of 8 (boson)+8 (fermion) component fields. 
To make clear the distinction between the two tensor multiplets, we 
call the former the `large (or, `massive') tensor multiplet' and 
the latter the `tensor gauge multiplet' (or, the `small tensor multiplet').

In this paper we assume knowledge of 5D superconformal 
tensor calculus. 
We refer the reader for the details to Ref.~\citen{ref:FO} (see also 
Ref.~\citen{ref:BCDWHP}).

The rest of this paper is organized as follows.
We start in \S2 by performing a dimensional reduction of the known
6D superconformal tensor multiplet to 5D, and find the 
small (8+8) tensor multiplet. The main purpose of that section is, however, 
to motivate the form of the supersymmetry transformation rule of 5D 
(large or small) tensor multiplets.  It is actually easier to compute 
directly in five dimensions for the purpose of determining the details of the 
multiplets. Therefore, in \S3 we examine directly in 5D the form of the 
supersymmetry transformation rule suggested by the dimensional 
reduction in a slightly more general manner, and actually find a 
larger 16+16 multiplet, which we call the `large tensor multiplet'. 
Here, we incorporate the point by Bergshoeff et~al.\cite{ref:BCWGHVP}
that the large tensor multiplets should be treated collectively with 
vector multiplets. In \S4, we impose a stronger constraint, which reduces 
the large tensor multiplet to a smaller 8+8 multiplet, and show that it 
coincides with the small (8+8) tensor multiplet obtained from 
the above mentioned dimensional reduction. 
In this case, the original tensor field $\BF_{ab}$ 
is made subject to the Bianchi-type constraint $\hat{\calD}_\nu\BF^{\mu\nu}
+\cdots=0$, and it is shown to be essentially expressed as a dual of the field 
strength $\partial_{[\lambda}\AB_{\mu\nu]}$ of a rank 2 tensor gauge field $\AB_{\mu\nu}
$. In \S5 we present the general form of the large tensor multiplet action. 
An explicit component form is given. 
It is shown that the mixing of the vector multiplets in the Yang-Mills gauge 
transformation of tne tensor multiplets can generally be eliminated aside 
from the mixing of the central-charge vector multiplet which is 
effectively induced by the vector-tensor mixing mass terms. Finally, in 
\S6 we briefly discuss the gauge invariant action for the small tensor 
multiplet. We also show there that the tensor gauge multiplet action is 
in fact dual to the vector multiplet action. An appendix is added to
present the solution of an algebraic equation that appears when
$\BF_{ab}$ is rewritten in
terms of the tensor gauge field $\AB_{\mu\nu}$.

\section{Dimensional reduction of a 6D tensor multiplet to 5D}


A superconformal tensor multiplet is known in 6D,\cite{ref:BSVP}  
but it is an on-shell multiplet;
that is, the superconformal algebra closes only when the 
multiplet satisfies an equation of motion.
We can, however, convert it into an off-shell 5D multiplet by making a 
dimensional reduction, in the same manner as we have done for the 
hypermultiplet.\cite{ref:KO1} 
When going down to 5D, we reinterpret the fifth spatial
derivative $\partial/\partial x^5\equiv\partial/\partial z$  as a {\em central charge transformation} $\XZ$, and 
then in 5D the constraint equations in 6D 
are no longer the equations of motion but become the defining 
equations of the central charge transformations of the relevant fields.
The curved index as well as the coordinate for the fifth spatial 
dimension is denoted $z$, in distinction from the tangent index 5; 
$(x^{\mu=0,1,2,3,4}, x^z=z)$.  
The $U(1)$ gauge multiplet $\XV^0=(M^0{\equiv}\alpha, W^0_\mu, {\Omega^0}^i, {Y^0}^{ij})$ 
which couples to this central charge is identified with 
the fifth spatial components of the vielbein, Rarita-Schwinger and 
conformal $SU(2)$ gauge fields,
$\6e_z{}^5,\ \6e_\mu{}^5,\ \6\psi_z^i,\ \6V_z^{ij}$,  
as follows (in the gauge $\6e_z{}^a=0$):\cite{ref:FO}
\begin{equation}
\alpha^{-1}=\6e_z{}^5,\quad W^0_{\mu}=\alpha\6{e}_{\mu}{}^5,
\quad \Omega_0^i=-\alpha^2\6{\psi}_z^i,
\quad Y_0^{ij}=\alpha^2\6{V}_z^{ij}-\frac{3i}{\alpha}\bar\Omega_0^i\Omega_0^j.
\end{equation}
The underline here denotes the fields in 6D. Note that 
$\6e_a{}^z=-\6e_a{}^\mu W_\mu^0$, so that the 6D derivative with 
flat index $\6{\partial}_a=\6e_a{}^\mu\partial_\mu+\6e_a{}^z\partial_z$ indeed reduces to 
the $\XZ$-covariant derivative 
$\6e_a{}^\mu(\partial_\mu-W_\mu^0\XZ)=\partial_a-W_a^0\XZ$ in 5D.

The 6D tensor multiplet
consists of a scalar $\6\sigma$, $SU(2)$-Majorana-Weyl 
spinor $\6\tau^i$\footnote{%
Originally, in Ref.~\citen{ref:BSVP}, this 
$SU(2)$-Majorana-Weyl spinor $\6\tau^i$ was denoted by $\psi^i$. We prefer 
$\tau^i$, since $\psi^i$ is easily confused with the Rarita-Schwinger field $
\psi_\mu^i$.}
 ($\gamma_7\6\tau^i=-\6\tau^i)$, and selfdual rank 3 tensor 
$F^+_{abc}$, whose superconformal transformation rules and constraint 
equations were given by Bergshoeff, Sezgin and Van Proeyen.\cite{ref:BSVP}
If we perform the dimensional reduction explained in Ref.~\citen{ref:KO1}
and use the 5D 
supersymmetry transformation, which is identified with a certain linear 
combination of the 6D superconformal transformations,\cite{ref:FO}
then we obtain a 5D tensor multiplet 
$(\sigma,\tau^i, \BF_{ab}, X^{ij},\cdots)$. 
The supersymmetry transformation laws of the first two components are 
found to be 
\begin{eqnarray}
\delta_Q(\varepsilon)\sigma&=& 2i\bar\varepsilon\tau,\nn
\delta_Q(\varepsilon)\tau^i &=& -\myfrac14\gamma^{ab}\BF_{ab}\varepsilon^i-\half\slashD\sigma\varepsilon^i
+\half\alpha\XZ\sigma\varepsilon^i+X^i{}_j\varepsilon^j,
\label{eq:DRtrf}
\end{eqnarray}
and the fields appearing here are identified with the following 
combinations of 6D fields:
\begin{eqnarray}
\sigma&=& {1\over\alpha}\6\sigma, \qquad 
\tau^i= {1\over\alpha}(\6\tau^i-\sigma\Omega_0), \nn
\BF_{ab}&=& {1\over\alpha}\left( F^+_{ab5}-2\alpha\sigma v_{ab} -\half\sigma\hat F_{ab}(W^0)
-2i\bar\Omega^0\gamma_{ab}\tau 
-\myfrac{i}{\alpha}\bar\Omega^0\gamma_{ab}\Omega^0 \right), \nn
X^{ij}&=& -{1\over\alpha}\left( \sigma Y_0^{ij} - 2i\bar\Omega^{0^(i}\tau^{j)} \right).
\label{eq:DRrel}
\end{eqnarray}
The powers of $\alpha$ have been multiplied such that $\sigma$, $\tau^i$ and 
$\BF_{ab}$ carry Weyl weights $w=1$, 3/2 and 2, respectively.

We could continue this procedure of dimensional reduction to find the 
transformation rule of the $\BF_{ab}$ field and to rewrite the constraint 
equations 
\begin{subeqnarray}
G_{\6a\6b}&\equiv&\hatD^{\6c}F^+_{\6a\6b\6c}+\cdots=0, 
\slabel{eq:Ca}\\
\Gamma^i&\equiv&\slashD\6\tau^i +\cdots=0, 
\slabel{eq:Cb}\\
C&\equiv&\hatD^{\6c}\hatD_{\6c}\6\sigma+\cdots=0 
\slabel{eq:Cc}
\end{subeqnarray}
in terms of these 5D fields. With this we would obtain an off-shell 
{\em tensor gauge multiplet} in 5D, since the $(\6a\6b)=(a5)$ 
component of the 
constraints (\ref{eq:Ca}) gives the Bianchi identity 
$\hatD^b\BF_{ab}+\cdots=0$, implying that $\BF_{\mu\nu}$ is the dual of the 
field strength $3\partial_{[\mu}A_{\nu\rho]}$ of a rank 2 tensor gauge field 
$\AB_{\mu\nu}$. The other $(ab)$ component of Eq.~(\ref{eq:Ca}) defines 
the central charge transformation of $B_{ab}$, 
$\partial_zF^+_{ab5}\sim\XZ \BF_{ab}$, and 
the constraints (\ref{eq:Cb}) and (\ref{eq:Cc}) define the 
central charge transformations $\XZ \tau^i$ and $\XZ (Z\sigma)$, 
respectively, of the 
fermion $\tau^i$ and the auxiliary scalar $Z\sigma\equiv\XZ\sigma$. 
Thus this tensor gauge multiplet consists of the 
8 boson components $1(\sigma)+6(A_{\mu\nu})+1(Z\sigma)$ and 
8 fermion components $\tau^i$. (Note that the rank $r$ antisymmetric tensor 
field in $d$ dimensions has ${}_{d-1}C_r=(d-1)!/r!(d-1-r)!$ off-shell 
degrees of freedom while the non-gauge tensor has 
${}_{d}C_r=d!/r!(d-r)!$ components.)

We find a concrete form of these results for this tensor gauge 
multiplet in \S4, working directly in 5D as a 
special case of the more general tensor multiplet, which we discuss in 
the next section.

\section{Large tensor multiplet $T^\alpha$}

Let $\XV^I=(M^I,\ W_\mu^I,\ \Omega^{Ii},\ Y^{Iij})$ ($I=0,1,2,\cdots$) be the 
vector multiplets\cite{ref:FO} of the system in which the 
zero-th multiplet 
$\XV^0$ denotes 
the $U(1)$ vector multiplet coupling to the central charge $\XZ$, 
and the other $\XV^I$ ($I\geq1$) 
are the Yang-Mills multiplets of a gauge group $G'$. We write 
$U(1)_Z\times G'=G$. 
We consider a set of scalar fields $\{ \sigma^\alpha\}_{\alpha=1,2,\cdots}$, 
which give a representation 
of the Yang-Mills group $G'$ and carry the central charge $\XZ$ as well.
Let us start with the superconformal transformation law $\delta\sigma^\alpha\equiv 
(\delta_Q(\varepsilon)+\delta_S(\eta))\sigma^\alpha=2i\bar\varepsilon^i\tau_i^\alpha\equiv2i\bar\varepsilon\tau^\alpha$ 
of the scalar fields $\sigma^\alpha$ with Weyl weight $w$. 
This defines the $SU(2)$-Majorana fermion field $\tau^i$. 
Then the 5D superconformal algebra 
presented in Ref.~\citen{ref:FO} 
generally determines the 
superconformal transformation of $\tau^i$ in 
the form
\begin{equation}
\delta\tau^i = -\myfrac14\gamma^{ab}\BF_{ab}\varepsilon^i-\myfrac12\slashD\sigma\varepsilon^i
+\myfrac12M_*\sigma\varepsilon^i-X^{ij}\varepsilon_j
-Z^{ij}_a\gamma^a\varepsilon_j-w\,\sigma\,\eta^i,
\label{eq:gtautrf}
\end{equation}
where $\BF_{ab}$ is an anti-symmetric tensor, and $X^{ij}$ and $Z^{ij}_a$ are 
an $SU(2)$-triplet [i.e., $(i,j)$-symmetric ] scalar and vector, respectively.
We use $\Lambda_*\varphi$ to denote the $G=U(1)_Z\times G'$ gauge transformation of 
the field $\varphi$ with parameters $\Lambda^I$:
\begin{equation}
\Lambda_*\sigma=\delta_G(\Lambda)\sigma=\Lambda^0\XZ \sigma+ \delta_{G'}(\Lambda)\sigma 
\end{equation}
We defer the presentation of the explicit form of the Yang-Mills 
$G'$-transformation $\delta_{G'}(\Lambda)\sigma$ of the tensor multiplet to 
\S{}\ref{sec:Action}, since the following discussion in this section is 
independent of it.
Comparing this general form (\ref{eq:gtautrf}) with the 
previous $\delta\tau^i$ given in Eq.~(\ref{eq:DRtrf}) for the tensor gauge multiplet, 
we see that the $SU(2)$-tensor 
vector term $Z^{ij}_a\gamma^a\varepsilon_j$ is missing in the latter. 
We are thus led to try the transformation rule (\ref{eq:gtautrf}) 
with the $Z^{ij}_a$ term {\em omitted}. Then,
using the 5D superconformal algebra,\cite{ref:FO} 
we find the transformation rules 
\begin{eqnarray}
\delta\sigma^\alpha&=&2i\bar\varepsilon\tau^\alpha,\nn
\delta\tau^{\alpha i}&=&
-\myfrac14\gamma^{ab}\BF_{ab}^\alpha\varepsilon^i
-\myfrac12\slashD \sigma^\alpha\varepsilon^i+\myfrac12M_*\sigma^\alpha\varepsilon^i
-X^{\alpha ij}\varepsilon_j-\sigma^\alpha\eta^i,\nn
\delta\BF^\alpha_{ab}&=&4i\bar\varepsilon\gamma_{[a}\hatD_{b]}\tau^\alpha 
-2i\bar\varepsilon\gamma_{cd[a}\gamma_{b]}\tau^\alpha v^{cd}
+2i\bar\varepsilon\hatR_{ab}(Q)\sigma^\alpha\nn
&&+2i\bar\varepsilon\gamma_{ab}\Omega_*\sigma^\alpha 
+2i\bar\varepsilon\gamma_{ab}M_*\tau^\alpha-
4i\bar\eta\gamma_{ab}\tau^\alpha,\nn
\delta X^{\alpha ij}&=&2i\bar\varepsilon^{(i}\slashD\tau^{\alpha j)}-i\bar\varepsilon^{(i}\gamma\dt v\tau^{\alpha j)}
-\myfrac{i}4\bar\varepsilon^{(i}\chi^{j)}\sigma^\alpha\nn
&&+4i\bar\varepsilon^{(i}\Omega^{j)}_*\sigma^\alpha+2i\bar\varepsilon^{(i}M_*\tau^{\alpha j)}
-2i\bar\eta^{(i}\tau^{\alpha j)}
\label{eq:Tensortrf}
\end{eqnarray}
where we have fixed the Weyl weight value of $\sigma$ to 1 
for convenience. 
(We can adjust it by multiplying by $\alpha=M^0$ if necessary.) 
A dot between two tensors generally represents
contraction; e.g., $\gamma\dt v=\gamma^{ab}v_{ab}$.  
%
Closure of the algebra requires the constraints  
\begin{eqnarray}
 0&=&M_*\BF^\alpha_{ab}+\hat F_{ab}(W)_*\sigma^\alpha 
+4v_{ab}M_*\sigma^\alpha+2i\bar\Omega_*\gamma_{ab}\tau^\alpha\nn
&&\qquad \qquad \qquad {}-\myfrac12\epsilon_{abcde}\left(
\hatD^c\BF^{\alpha de}+2i\bar\tau^\alpha\gamma^c\hatR^{de}(Q)\right),\label{eq:ZF}\\ 
0&=&M_*X^{\alpha ij}+Y^{ij}_*\sigma^\alpha-2i\bar\Omega^{(i}_*\tau^\alpha{}_{\Span]}^{j)}
\label{eq:ZX}
\end{eqnarray}
and the supersymmetry transformation descendants of  
(\ref{eq:ZX}), since it is not $\XQ$-inert. 
Actually, we now see that the quantity on the RHS of Eq.~(\ref{eq:ZX}) turns 
out to be the first component of a {\em linear multiplet}.\cite{ref:FO}

We should observe that this transformation rule (\ref{eq:Tensortrf})
for the tensor multiplet $\XT^\alpha$ 
has exactly the same form as that\cite{ref:FO} 
for the vector multiplet $\XV^I$. 
The only difference is hidden in the $G$-transformation `${}_*$' and 
its manifestations is that the vector multiplets carry no central charge, $\XZ 
\XV^I=0$. Then, it is easy to see that all the constraints in (\ref{eq:ZF}) 
and (\ref{eq:ZX}) are trivially satisfied when the component fields of
the tensor multiplets $\XT^\alpha=(\sigma^\alpha,\,\tau^{\alpha i},\,\BF_{ab}^\alpha,\,X^{\alpha ij})$ are
replaced by those of vector multiplets, $\XV^I=(M^I,\,\Omega^{Ii},\hat 
F_{ab}^I,\,Y^{Iij})$.  
In any case, since the transformation rules take 
the same form, the embedding formula of the vector multiplets $V^I$ into a 
linear multiplet $\XL(f(V))$,\cite{ref:FO} 
which applies to any homogeneous quadratic 
function $f(V)=\half f_{IJ} V^IV^J$, is valid even when the tensor 
multiplets $T^\alpha$ are included, and the formula is generalized as 
follows. We write the vector and tensor multiplets collectively as 
$\calT^A\equiv(V^I,\,T^\alpha)$, and their components as $\calT^A=(\sigma^A,\,\tau 
^{Ai},\,\BF_{ab}^A,\,X^{Aij})$, following the notation for the tensor 
multiplet. For the vector multiplet index $A=I$, of course, 
the components should be 
understood as $(\sigma^I,\,\tau^{Ii},\,\BF_{ab}^I,\,X^{Iij}) = (M^I,\,\Omega 
^{Ii},\hat F_{ab}^I(W),\,Y^{Iij})$. 
Then, for any quadratic function 
$f(\calT)=\myfrac12f_{\etI \etJ }\calT^\etI \calT^\etJ =
\myfrac12f_{IJ}V^IV^J+f_{I\alpha}V^IT^\alpha+
\myfrac12f_{\alpha\beta}T^\alpha T^\beta$, 
we have the following linear multiplet 
$\XL(f(\calT))=(\,L^{ij}(f(\calT)),\, \varphi^i(f(\calT)),\,
E_a(f(\calT))$,\, $N(f(\calT))\,)$: 
%
\begin{eqnarray}
L^{ij}(f(\calT))&=&f_\etI  X^{\etI ij}-i\bar\tau^{\etI i}\tau^{\etJ j}f_{\etI \etJ }, \nn 
\varphi^i(f(\calT))&=&
-\myfrac14\chi^if+2\Omega^i_*f
+\left(\slashD-\myfrac12\gamma\dt v+M_*\right)\tau^{\etI i}f_\etI \nn
&&{}+\left(-\myfrac14\gamma\dt \BF^\etI 
+\myfrac12\slashD \sigma^\etI -\myfrac12M_*\sigma^\etI -X^\etI \right)
\tau^{\etJ i}f_{\etI \etJ },\nn
E_a(f(\calT))&=&\hatD^b\left(4v_{ab}f+\BF^\etI _{ab}f_\etI 
+i\bar\tau^\etI \gamma_{ab}\tau^\etJ f_{\etI \etJ }\right)
+\myfrac18\epsilon_{abcde}\BF^{\etI bc}\BF^{\etJ de}f_{\etI \etJ }\nn
&&{}+\left(-\hatD_aM_*\sigma^\etI +\hatD_a\sigma^\etI M_*\right)f_\etI \nn
&&{}+\left(-2i(\bar\Omega_*\gamma_a\tau)^\etI \sigma^\etJ +2i\bar\tau^\etI \gamma^a(M_*\tau)^\etJ 
+4i\bar\tau^\etI \gamma_a(\Omega_*\sigma)^\etJ \right)f_{\etI \etJ },\nn
N(f(\calT))&=&-\hatD^a\hatD_af+M_*M_*f-(\myfrac12D+3v\dt v)f
+(-2\BF^\etI \dt v+i\bar\chi\tau^\etI +2i\bar\Omega_*\tau^\etI )f_\etI \nn
&&{}+\left(-\myfrac14\BF^\etI \dt \BF^\etJ +\myfrac12\hatD^a\sigma^\etI \hatD_a\sigma^\etJ 
+2i\bar\tau^\etI \slashD \tau^\etJ -i\bar\tau^\etI \gamma\dt v\tau^\etJ 
+X^\etI _{ij}X^{\etJ ij}\right.\nn
&&\qquad \qquad {}\left.-\myfrac32M_*\sigma^\etI M_*\sigma^\etJ 
+4i\bar\tau^\etI M_*\tau^\etJ +4i\bar\tau^\etI \Omega_*\sigma^\etJ \right)
f_{\etI \etJ },
\label{eq:ZT}
\end{eqnarray}
where 
\begin{eqnarray}
&&f\equiv f(\sigma)=\myfrac12f_{\etI \etJ }\sigma^\etI \sigma^\etJ =
\myfrac12f_{IJ}M^IM^J+f_{I\alpha}M^I\sigma^\alpha+
\myfrac12f_{\alpha\beta}\sigma^\alpha\sigma^\beta, \nn
&&f_{\etI }={\partial f(\sigma)\over\partial\sigma^{\etI }}, \qquad  
f_{\etI \etJ }={\partial^2f(\sigma)\over\partial\sigma^{\etI }\partial\sigma^{\etJ }}
\end{eqnarray}
in this formula.

In view of the formula (\ref{eq:ZT}), we recognize that the RHS 
quantity of the constraint (\ref{eq:ZX}) is just the 
first component 
$L^{ij}(f(\calT))$ of the linear multiplet $L(f(\calT))$ for the choice $f(\calT)=V_*T^\alpha$:
\begin{equation}
f(\sigma)=M_*\sigma^\alpha= M^0\XZ \sigma^\alpha+\delta_{G'}(M)\sigma^\alpha.
\end{equation}
Thus, the complete set of constraint equations for the tensor multiplets 
$T^\alpha$ are given by Eq.~(\ref{eq:ZF}) and 
\begin{equation}
\XL(V_*T^\alpha)=\left(L^{ij}(V_*T^\alpha),\, \varphi^i(V_*T^\alpha),
\,E_a(V_*T^\alpha),\,N(V_*T^\alpha)\right) =0.
\label{eq:Tconstrt}
\end{equation}
It is, however, easy to see that 
the vector component constraint $E_a(V_*T)=0$ here
is automatically satisfied if the 
constraint (\ref{eq:ZF}) is satisfied. In confirming this, we note that 
the last two lines of  $E_a(f(\calT))$ in Eq.~(\ref{eq:ZT}) vanish for 
$f(\calT)=V\dt T$, with the dot product `$\cdot$' satisfying 
Eq.~(\ref{eq:Jacobilike}) in the footnote appearing subsequently, 
and thus in particular, for the simple product $VT$ or 
$*$-product $V_*T$.
We also need the equation
\begin{eqnarray}
\hatD_{[a}\BF^\alpha_{bc]}+2i\bar\tau^\alpha\gamma_{[a}\hatR_{bc]}(Q)&=&
e_{[a}{}^\mu e_b{}^\nu\calD_{c]}\cBF^\alpha_{\mu\nu} 
-2i\bar\psi_{[a}\gamma_{bc]}\Omega_*\sigma^\alpha\nn
&&{}-2i\bar\psi_{[a}\gamma_{bc]}M_*\tau^\alpha 
+2i\bar\psi_{[a}\gamma_b\psi_{c]}M_*\sigma^\alpha,
\hspace{2em}\label{eq:RQidentity}
\end{eqnarray}
where we have introduced
$\cBF_{\mu\nu}^A=\{\cBF_{\mu\nu}^\alpha,F_{\mu\nu}^I(W)\}$ without a hat, 
`\ $\hat{}$\ ' 
,representing $\BF_{\mu\nu}^A$ 
with supersymmetry covariantization terms subtracted:
\begin{equation}
\cBF^A_{ab}\equiv\BF^A_{ab}-4i\bar\psi_{[a}\gamma_{b]}\tau^A
+2i\bar\psi_a\psi_c\sigma^A.
\end{equation}
Equation (\ref{eq:RQidentity}) can be shown by using explicit expressions for 
$\hatR_{ab}(Q)$ and supercovariant derivatives as well as the identity
$\gamma_d\psi^i_{[a}(\bar\psi^{\Span{[}}_b\gamma^d\psi_{c]})
=\psi^i_{[a}(\bar\psi^{\Span{[}}_b\psi_{c]})$ 
[See Eq.~(A$\cdot$11) in Ref.~\citen{ref:KO1}]. 
Thus the independent constraints 
are given by Eq.~(\ref{eq:ZF}), $L^{ij}(V_*T)=0, \varphi^i(V_*T)=0$
and $N(V_*T)=0$, which are interpreted as defining equations of the 
central charge transformation of $\BF_{ab},\ X^{ij},\ Z\tau^i$ 
and $\ Z^2\sigma$, 
respectively. ($Z^n\phi$ represents the field $\XZ^n\phi$ obtained by 
performing central charge transformation $n$ times.)
We thus finally see that this tensor multiplet 
consists of 
\begin{equation}
T^\alpha=\bigl(\sigma^\alpha,\ \tau^{\alpha i},\ \BF^\alpha_{ab},\ X^{\alpha ij},\ Z\sigma^\alpha,\ 
Z\tau^{\alpha i},\ Z^2\sigma^\alpha\bigr)
\end{equation}
for each $\alpha$ and contains 
$1(\sigma)+10(\BF_{ab})+3(X^{ij})+1(Z\sigma)+1(Z^2\sigma)=16$ bose field components 
and $8(\tau^i)+8(Z\tau^i)=16$ fermi field components.
Note that $\BF_{ab}$ is a mere tensor field, neither a field strength 
nor a gauge potential field.  We call this 16+16 multiplet 
the `large tensor multiplet', contrasting it with the smaller `tensor gauge 
multiplet'. The components of this multiplet and their properties are
listed in Table \ref{table:1}.

The superconformal transform action rules for the auxiliary fields 
$Z\sigma^\alpha,\ Z\tau^{\alpha i}$ and $Z^2\sigma^\alpha$ are omitted, 
since they trivially follow from the stipulation that the central charge 
transformation $\XZ$ be {\em central}, i.e., that they commute with all 
transformations. 
\begin{table}[tb]
\caption{Field content of the multiplets.}
\label{table:1}
\begin{center}
\begin{tabular}{lcccc} \hline\hline
  field  &type & restrictions & $SU(2)$ &Weyl-weight       \\ \hline 
  \multicolumn{5}{c}{large tensor multiplet ${\mbf T}$}   \\ \hline
$\sigma$  &boson  &  real&{\mbf 1} &  $1$    \\  
$\tau^i$&fermion  &$SU(2)$-Majorana &{\mbf 2}&   $\myfrac32$   \\  
$\hat B_{ab}$&boson& real, antisymmetric  &{\mbf 1}& $2$  \\  
$X^{ij}$    &boson  &$X^{ij}=X^{ji}=(X_{ij})^*$ &{\mbf 3}& $2$    \\ 
$Z\sigma$     & boson & real  &{\mbf 1} &$1$\\ 
$Z\tau^i$   & fermion& $SU(2)$-Majorana  &{\mbf 2} &$\myfrac32$\\  
$Z^2\sigma$   & boson  &  real         &{\mbf 1}        &$1$ \\ \hline
  \multicolumn{5}{c}{tensor gauge multiplet ${\mbf A}$}   \\ \hline
$A_{\mu\nu}$&boson&real,$~\delta A_{\mu\nu}=2\partial_{[\mu}\Lambda_{\nu]}$&{\mbf 1}& $0$  \\  
$\sigma$  &boson  & real &{\mbf 1} &  $1$    \\  
$\tau^i$&fermion  &$SU(2)$-Majorana&{\mbf 2}&   $\myfrac32$   \\  
$Z\sigma$     & boson &real   &{\mbf 1} &$1$\\ \hline
\end{tabular}
\end{center}
\end{table}

\section{Tensor gauge multiplet $\GT$}

If the tensor multiplet carries 
no gauge group $G'$ charges other than the central charge, 
that is, $V_*T=V^0\XZ T$, 
then we can derive a smaller tensor multiplet by requiring
\begin{equation}
\XL(V^0T)=\left(L^{ij}(V^0T),\, \varphi^i(V^0T),
\,E_a(V^0T),\,N(V^0T)\right) =0,
\label{eq:TGconstrt}
\end{equation}
which is a stronger constraint than the previous one, 
$\XL(V_*T)=0$, in Eq.~(\ref{eq:Tconstrt}). Indeed, the latter follows 
from the former; $\XL(V_*T)=\XZ\XL(V^0T)=0$, so that the constraint 
(\ref{eq:TGconstrt}) plus Eq.~(\ref{eq:ZF}) gives a sufficient set 
of conditions for the multiplet to exist. Note that the first component 
of this linear multiplet constraint (\ref{eq:TGconstrt}) reads
\begin{equation}
L^{ij}(V^0T)=\alpha X^{ij}+\sigma Y_0^{ij}-2i\bar\Omega_0^{(i}\tau_{\Span{]}}^{j)}=0,
\label{eq:TGconstrt1}
\end{equation}
which can be solved with respect to $X^{ij}$ so that 
$X^{ij}$ is no longer an independent field! 
This contrasts with the case of the large tensor multiplet,in which the 
constraint $L^{ij}(V_*T)=0$ merely defines the central charge 
transformation of $X^{ij}$, $\XZ X^{ij}$. 
We here notice that 
Eq.~(\ref{eq:TGconstrt1}) is the same relation as 
the last equation in Eq.~(\ref{eq:DRrel}) that we obtained by 
dimensional reduction from the 6D tensor multiplet. We have therefore found 
that the relation (\ref{eq:DRrel}) is the first component of the 
linear multiplet constraint (\ref{eq:TGconstrt}), and hence that 
the multiplet obtained by imposing the stronger constraints 
$L(V^0T)=0$, (\ref{eq:TGconstrt}), is nothing but the {\em 
tensor gauge multiplet} possessing 8 boson + 8 fermion components 
that we would get by dimensional reduction from the 6D tensor multiplet. 

Note that the constraints $\varphi^i(V^0T)=0$ and $N(V^0T)=0$ 
determine the central charge transformations $\XZ \tau^i$ and $\XZ (Z\sigma)$, 
respectively. The additional constraint (\ref{eq:ZF}) determines the 
the central charge transformation of $\BF_{ab}$. As we now show in detail, 
the final constraint $E_a(V^0T)=0$ corresponds to the 
(a5) components of the 6D constraints (\ref{eq:Ca}) and is 
the Bianchi identity $\hatD^b\BF_{ab}+\cdots=0$.\footnote{%
Actually, this tensor gauge multiplet can be defined even if 
we replace the linear multiplet constraint (\ref{eq:TGconstrt1}), 
$L(V^0T)=0$, 
defined with a simple product of $V^0$ and $T$, by a more general 
one, $L(V\dt T)=0$ defined with a {\em dot product} ` $\cdot$ ' 
 that satisfies the Jacobi-like identity
\begin{eqnarray}
[V_1,\,V_2;\, T]\equiv 
V_1\cdot({V_2}_*T)-V_2\cdot({V_1}_*T)-({V_1}_*{V_2})\cdot T
=0
\label{eq:Jacobilike}
\end{eqnarray}
and invertibility. 
Then $L(V\dt T)=0$ implies $L(V_*T)=0$, since 
\begin{eqnarray}
\delta_G(\Lambda)\XL(V\dt T)=\Lambda_*\XL(V\dt T)
=\XL((\Lambda_*V)\dt T+V\dt (\Lambda_*T))
=\Lambda\dt \XL((V_*T))-\XL([\Lambda,V;T]).
\end{eqnarray}
The Bianchi identity $E_a(V\dt T)=0$ can also be solved with a rank 2
tensor gauge field $\AB_{\mu\nu}$ in the same way as for the simple product 
case $E_a(V^0T)=0$. In the main text, however, we deal with only the latter 
case for simplicity of notation. }

Generally, the linear multiplet $(L^{ij}, \varphi^i, E_a, N)$ satisfies 
the constraint 
\begin{equation}
\hatD_aE^a+i\bar\varphi\gamma\dt \hatR(Q)+M_*N+4i\bar\Omega_*\varphi
+2Y^{ij}_*L_{ij}=0,
\end{equation} 
which can be rewritten in the form\cite{ref:FO}
\begin{eqnarray}
&&e^{-1}\partial_\lambda(e
{\cal E}^\lambda)+2{\cal H}_{VL}=0,\label{eq:Con.V} 
\label{eq:diveq} \\
&&\hbox{with} \quad 
{\cal E}^\lambda\equiv E^\lambda-2i\bar\psi_\rho\gamma^{\rho\lambda}\varphi
+2i\bar\psi_\rho\gamma^{\lambda\rho\sigma}L\psi_\sigma,\nn
&&\hspace*{2em}
{\cal H}_{VL}\equiv Y^{ij}_*L_{ij}+2i\bar\Omega_*\varphi
 +2i\bar\psi^a_i\gamma_a\Omega_{j*}L^{ij}-\myfrac12W_{a*}{\cal E}^a \nn
&&\hspace*{6.3em}{}+\myfrac12M_*\left(N-2i\bar\psi_b\gamma^{b}\varphi
-2i\bar\psi_a^{(i}\gamma^{ab}\psi_b^{j)}L_{ij}\right).\label{eq:def.VH}
\end{eqnarray}
In the present case of the linear multiplet $L(V^0T)$, it is $G'$-neutral 
and the $*$-operation is only the $\XZ$ transformation. If we here 
use the constraints 
$L^{ij}(V^0T)=\varphi^i(V^0T)=N(V^0T)=0$ other than $E^\lambda(V^0T)=0$, 
this equation (\ref{eq:diveq}) is reduced to 
\begin{equation}
e^{-1}\partial_\lambda(e E^\lambda(V^0T))- W^0_\lambda\XZ E^\lambda(V^0T) =0.
\end{equation}
But $\XZ E^\lambda(V^0T)=E^\lambda(V_*T)$, which vanishes
automatically because of the 
constraint (\ref{eq:ZF}), as remarked before. Thus we have 
$\partial_\lambda(e E^\lambda(V^0T))=0$, just as in the case when $L(V^0T)$ is 
completely neutral.  This implies that $e E^\lambda(V^0T)$ can be 
written as the divergence of a rank 2 antisymmetric tensor density 
$E^{\lambda\rho}$. Indeed, 
inspecting the formula (\ref{eq:ZT}) for $f(M)=\alpha\sigma$, we can show that 
$E_a(V^0T)$ can be written in the form
%
%
\begin{eqnarray}
E^\lambda(V^0T)&=& -e^{-1}\partial_\rho( E^{\lambda\rho}(V^0T)), 
\label{eq:form} \\
-E^{\lambda\rho}(V^0T)&=&
e\left(4v^{\lambda\rho}\alpha\sigma+\alpha\BF^{\lambda\rho}+\hat F^{\lambda\rho}(W^0)\sigma 
+2i\bar\Omega^{0}\gamma^{\lambda\rho}\tau\right) \nn
&&\hspace{-1em}{}	+\myfrac12\epsilon^{\lambda\rho\sigma\mu\nu}
\bigl\{W^0_\sigma\cBF_{\mu\nu} 
+2i\bar\psi_\mu\gamma_\sigma\psi_\nu\alpha\sigma 
+2i\bar\psi_\sigma\gamma_{\mu\nu}(\alpha\tau+\Omega^0\sigma)
\bigr\}.\hspace{2em}
\end{eqnarray}
This expression actually has the same form as 
$E^{\lambda\rho}(V_1V_2)$\cite{ref:FO} 
for the case of the completely neutral vectors $V_1$ and $V_2$. 

Now, because of the form (\ref{eq:form}), 
the constraint $E^\lambda(V^0T)=0$ 
implies that there exists an 
antisymmetric tensor gauge field $\AB_{\mu\nu}$ with which 
$-E^{\lambda\rho}$ can be written as 
$\myfrac1{3!}\epsilon^{\lambda\rho\sigma\mu\nu}
\partial_{[\sigma}\AB_{\mu\nu]}$; thus the constraint $E^\lambda(V^0T)=0$ is
rewritten as
\begin{equation}
\hat F_{\lambda\mu\nu}(\AB)-\myfrac12\epsilon_{\lambda\mu\nu\rho\sigma}
\left(4v^{\rho\sigma}\alpha\sigma+\alpha\BF^{\rho\sigma}+\hat F^{\rho\sigma}(W^0)\sigma 
+2i\bar\Omega^{0}\gamma^{\rho\sigma}\tau\right)=0
\label{eq:Econstrt}
\end{equation}
in terms of the covariant field strength of $\AB_{\mu\nu}$:
\begin{eqnarray}
\hat F_{\lambda\mu\nu}(\AB)&\equiv&
3\partial_{[\lambda}\AB_{\mu\nu]}
-3W^0_{[\lambda}\cBF_{\mu\nu]}\nn
&&{}-6i\bar\psi_{[\lambda}\gamma_{\mu\nu]}(\alpha\tau+\Omega^0\sigma)
+6i\bar\psi_{[\lambda}\gamma_\mu\psi_{\nu]}\alpha\sigma\,.
\label{eq:Bstrength}
\end{eqnarray}
With Eq.~(\ref{eq:Econstrt}), the original tensor 
$\BF_{ab}$ is now rewritten in terms of the {\em tensor gauge field} 
$\AB_{\mu\nu}$, which has fewer components, 
${}_{5-1}C_2=6$, than $\BF_{ab}$,  
due to gauge invariance under
$\delta\AB_{\mu\nu}=2\partial_{[\mu}\Lambda^B_{\nu]}$.
Solving $\BF_{ab}$ in terms of $\AB_{\mu\nu}$ is, however, not quite trivial, 
since Eq.~(\ref{eq:Econstrt}) contains $\BF_{ab}$ in two places; that is, 
it has the form 
\begin{eqnarray}
&&3W^0_{[\lambda}\BF_{\mu\nu]}+
\myfrac12\alpha\epsilon_{\lambda\mu\nu\rho\sigma}\BF^{\rho\sigma} = \FA _{\lambda\mu\nu}, 
\label{eq:bianchi} \\
&&\FA _{\lambda\mu\nu}\equiv 
3\partial_{[\lambda}\AB_{\mu\nu]}
+6iW^0_{[\lambda}(2\bar\psi_\mu\gamma_{\nu]}\tau-\bar\psi_\mu\psi_{\nu]}\sigma)
-6i\bar\psi_{[\lambda}\gamma_{\mu\nu]}(\alpha\tau+\Omega^0\sigma) \nn
&&\qquad{}+6i\bar\psi_{[\lambda}\gamma_\mu\psi_{\nu]}\alpha\sigma 
-\myfrac12\epsilon_{\lambda\mu\nu\rho\sigma}
\left(4v^{\rho\sigma}\alpha\sigma+\hat F^{\rho\sigma}(W^0)\sigma 
+2i\bar\Omega^{0}\gamma^{\rho\sigma}\tau\right).
\label{eq:delta}
\end{eqnarray}
This is solved in the Appendix to yield
\begin{equation}
\BF_{ab}={\alpha\over\alpha^2-(W^0)^2}\left(
\frac1{3!}\epsilon_{abcde}\FA ^{cde} 
-\frac1{\alpha}W^{0c}\FA _{abc}
+\frac2{3!\alpha^2}W^0_{[a}\epsilon^{\Span{]}}_{b]cdef}W^{0c}\FA ^{def}
\right). \label{eq:sol}
\end{equation}

The transformation law of $\AB_{\mu\nu}$ can be read from this covariant 
field strength (\ref{eq:Bstrength}) as
\begin{eqnarray}
\delta\AB_{\mu\nu}&=&
2i\bar\varepsilon\gamma_{\mu\nu}(\alpha\tau+\Omega^0\sigma)-4i\bar\varepsilon\gamma_{[\mu}\psi_{\nu]}\alpha\sigma 
+W_{[\mu}^0 (4i\bar\varepsilon\gamma_{\nu]}\tau-4i\bar\varepsilon\psi_{\nu]}\sigma)\nn
&&{}+2\partial_{[\mu}\Lambda^B_{\nu]}+\Lambda^0\cBF_{\mu\nu} ,
\end{eqnarray}
where $\Lambda^0$ is the parameter of the central charge transformation $\XZ$. 
Thus, the independent components of the tensor gauge multiplet are 
\begin{equation}
\GT = \bigl(\sigma,\  \tau^i,\ \AB_{\mu\nu},\  Z\sigma\bigr),
\end{equation}
and their properties are listed in Table \ref{table:1}.

We add here the transformation law of the field strength 
$\hat F_{\mu\nu\rho}(\AB)$ and its Bianchi identity 
(equivalent to $E_a(V^0T)=0$):
\begin{eqnarray}
\delta\hat F_{abc}(\AB)&=&6i\bar\varepsilon\gamma_{[ab}\hatD_{c]}(\alpha\tau+\Omega^0\sigma)
+6i\bar\varepsilon\gamma_{[a}\hatR_{bc]}(Q)\alpha\sigma\nn
&&{}+3i\bar\varepsilon\gamma_{de[a}\gamma_{bc]}(\alpha\tau+\Omega^0\sigma)v^{de} 
+6i\bar\varepsilon\gamma_{[a}(\hat F_{bc]}(W^0)\tau+\Omega^0\BF_{bc]}) \nn
&&{}+6i\bar\eta\gamma_{abc}(\alpha\tau+\Omega^0\sigma)
+3\Lambda^0(\hatD_{[a}\BF_{bc]}+2i\bar\tau\gamma_{[a}\hatR_{bc]}(Q)) \,, \nn
0&=&\hatD_{[a}\hat F_{bcd]}(\AB)
+\myfrac34 \BF_{[ab}^\etI \BF_{cd]}^\etJ f_{\etI \etJ }\ .
\end{eqnarray}

\section{%
An invariant action for large tensor multiplets
\label{sec:Action} }

We first discuss the explicit form of the Yang-Mills $G'$-gauge 
transformation of the large tensor multiplets $T^\alpha$. 
We can interpret the vector multiplets as special tensor multiplets 
that are $\XZ$-inert. This fact enabled us to treat 
the tensor and vector multiplets collectively as $\calT^A\equiv(V^I,\,T^\alpha)$
in the embedding formula $\XL(f(\calT))$ given in Eq.~(\ref{eq:ZT}). 
This suggests that the tensor and vector multiplets transform 
collectively also under the $G'$-gauge transformation; that is, 
$\delta_{G'}(\Lambda)\calT^A= \Lambda^J(t_J)^A{}_B\calT^B$, or more explicitly, 
\begin{equation}
\delta_{G'}(\Lambda)\pmatrix{ V^I \cr T^\alpha\cr}= \sum_{J\geq1}\Lambda^J 
\pmatrix{ 
(t_J)^I{}_K & 0 \cr 
(t_J)^\alpha{}_K & (t_J)^\alpha{}_\beta\cr}
\pmatrix{ V^K \cr T^\beta\cr}.
\label{eq:Gtrf0}
\end{equation}
Here, we have used 0 for the top-right entry of the generator matrix 
$(t_J)^A{}_B$, because 
the tensor multiplets are $\XZ$-variant and therefore cannot appear 
in the $G'$-gauge transformation 
of $\XZ$-inert vector multiplets $V^I$. 
The other off-diagonal entry, $(t_J)^\alpha{}_K$, can be
non-vanishing. Specifically, 
the $G'$-gauge transformation of the scalar components 
$\sigma^\alpha$ of the tensor multiplets, $\delta_{G'}(\Lambda)\sigma^\alpha$, can generally 
contain the scalar fields $M^I$ of a vector multiplet $\XV^I$ as well, as 
 pointed out very recently by Bergshoeff et~al.\cite{ref:BCWGHVP}
Nevertheless, as we see shortly, this mixing of vector multiplets 
in the $G'$-transformation of the tensor multiplets is only apparent, and it
must vanish in the field basis in which the kinetic terms of tensor and 
vector multiplets do not mix with each other: $(t_I)^\alpha{}_K=0$ for
$I\geq1$. 
We see below, however, that the introduction of a mass term effectively induces
an off-diagonal entry $(t_0)^\alpha{}_K$ only for the central charge
transformation, $I=0$.  

A general invariant action for the tensor multiplets 
is obtained by using the VL action formula\cite{ref:FO} as follows:
\begin{eqnarray}
{\cal L}_T&=&{\cal L}_{\rm VL}\left(V^0 \XL(h(\calT))\right),\qquad 
h(T)= -\calT^A(\XZ \calT^B)d_{AB}+\calT^A\calT^B\eta_{AB}.
\label{eq:preTaction}
\end{eqnarray}   
Here we have put a negative sign in front of the kinetic term for
later convenience. Because the linear multiplet $\XL(h(\calT))$ has
a non-zero central charge, $V^0$ in Eq.~(\ref{eq:preTaction}) must be the
central charge vector multiplet in order for the action to be
$\XZ$-invariant.
Because $\XZ V^I=0$, we can take $d_{AJ}=(d_{\alpha J},d_{IJ})=0$. We can also 
assume that the submatrix $d_{\alpha\beta}$ is invertible, because 
$T^\alpha(\XZ T^\beta)d_{\alpha\beta}$ gives the kinetic term of the tensor 
multiplets, as we see below. 
Then, we can redefine the tensor multiplets as
$T^\alpha\,\rightarrow\,T^\alpha-V^Id_{I\beta}(d^{-1})^{\beta\alpha}$ without changing the 
superconformal transformation rule (\ref{eq:Tensortrf}) to cancel the 
off-diagonal part of the kinetic term, $V^I(ZT)^\alpha d_{I\alpha}$. Further, 
the contribution of $V^IV^J\eta_{IJ}$ in $\calT^A\calT^B\eta_{AB}$ 
can be absorbed into the kinetic term of the vector multiplets, 
${\cal L}_{\rm VL}\bigl(V^I \XL(V^JV^K)\bigr)c_{IJK}$, 
which we also discuss later. 
The function $h(\calT)$ can, therefore, generally be assumed to be  
\begin{eqnarray}
h(\calT)&=&-T^\alpha(\XZ T)^\beta d_{\alpha\beta}+T^\alpha T^\beta\eta_{\alpha\beta}+2T^\alpha V^I\eta_{\alpha I}.
\label{eq:hT}
\end{eqnarray}
We can also assume without loss of generality that the metric tensor 
$d_{\alpha\beta}$ is antisymmetric, $d_{\alpha\beta}=-d_{\beta\alpha}$ (and $\eta_{\alpha\beta}$ is 
symmetric, $\eta_{\alpha\beta}=\eta_{\beta\alpha}$). This is because the symmetric 
part $d^S_{\alpha\beta}$ of $d_{\alpha\beta}$, if it exists, yields a total central-charge 
transformed linear multiplet $\XZ \XL(T^\alpha T^\beta)d^S_{\alpha\beta}$, but 
the action of the form ${\cal L}_{\rm VL}\left(V_0 \XZ\XL\right)$ is seen to 
vanish up to total derivative terms.

We examine the invariance of the action (\ref{eq:preTaction}) with 
$h(\calT)$ in Eq.~(\ref{eq:hT}) under the $G'$-transformation 
(\ref{eq:Gtrf0}). It is easily seen that $h(\calT)$ itself must be 
$G'$-invariant and that the kinetic term part 
$T^\alpha(\XZ T)^\beta d_{\alpha\beta}$ and the mass term part 
$T^\alpha T^\beta\eta_{\alpha\beta}+2T^\alpha V^I\eta_{\alpha I}$ must be 
separately invariant. The $G'$-transformation of the former gives
\begin{eqnarray}
\delta_{G'}(\Lambda)\bigl(T^\alpha(\XZ T)^\beta d_{\alpha\beta}\bigr)
&=&\Lambda^J(t_J)^\alpha{}_KV^K(\XZ T)^\beta d_{\alpha\beta} \nn
&&{}+\Lambda^J\bigl((t_J)^\gamma{}_\alpha d_{\gamma\beta}
+d_{\alpha\gamma}(t_J)^\gamma{}_\beta\bigr)\bigl(T^\alpha(\XZ T)^\beta\bigr),
\end{eqnarray}
and therefore it is necessary that 
the antisymmetric tensor $d_{\alpha\beta}$ is $G'$ invariant,
\begin{equation}
(t_J)^\gamma{}_\alpha d_{\gamma\beta}
+d_{\alpha\gamma}(t_J)^\gamma{}_\beta=0, \qquad (J=1,2,\cdots)
\end{equation}
and the off-diagonal entry $(t_J)^\alpha{}_K$ vanishes,
\begin{equation}
(t_J)^\alpha{}_K=0, \qquad (J=1,2,\cdots)
\end{equation}
as stated above. 
Similarly, the invariance of the mass term 
$T^\alpha T^\beta\eta_{\alpha\beta}+2T^\alpha V^I\eta_{\alpha I}$ 
requires the $G'$-invariance of the symmetric tensors 
$\eta_{\alpha\beta}$ and $\eta_{\alpha I}$:
\begin{equation}
(t_J)^\gamma{}_\alpha\eta_{\gamma\beta}+\eta_{\alpha\gamma}(t_J)^\gamma{}_\beta=0, \qquad 
(t_J)^\gamma{}_\alpha\eta_{\gamma I}+\eta_{\alpha K}(t_J)^K{}_I=0. 
 \qquad (J=1,2,\cdots)
\label{eq:EtaInv}
\end{equation}

We now wish to obtain an explicit component expression of the action 
(\ref{eq:preTaction}) with $h(\calT)$ in Eq.~(\ref{eq:hT}). For that 
purpose, we first compute the expression for the following simpler 
action without mass terms
\begin{eqnarray}
{\cal L}_T&=&-{\cal L}_{\rm VL}\left(V^0 \XL(T^\alpha(\XZ T)^\beta)\right)d_{\alpha\beta}.
\label{eq:Taction}
\end{eqnarray} 
As we show below, the expression for the general action with mass 
terms can easily be obtained from the result in this case.

The component expression of the action (\ref{eq:Taction}) is computed in
the following way. The constraints (\ref{eq:Tconstrt}) give complicated 
expressions for the central-charge transformed quantities $\XZ \BF^\alpha 
_{ab},\,\XZ X^{\alpha ij},\, 
\XZ(Z\tau^{\alpha i}),\,\XZ(Z^2\sigma^\alpha)$, which should be expressed in terms of the 
independent fields $\sigma^\alpha,\,\tau^{\alpha i},\,\BF^\alpha_{ab},\,X^{\alpha ij},\, 
Z\tau^{\alpha i}$, $Z\sigma^\alpha$ and $Z^2\sigma^\alpha$. However, fortunately, this can be done 
relatively easily as follows. Because
\begin{equation}
V_*T^\alpha= V^0\XZ T^\alpha+ gV^\alpha{}_\beta T^\beta \qquad 
\hbox{with} \quad  
V^\alpha{}_\beta= \sum_{I\geq1}V^I(t_I)^\alpha{}_\beta,
\end{equation}
the constraints (\ref{eq:Tconstrt}), $\XL(V_*T^\alpha)=0$, can be 
rewritten in the form
\begin{equation}
\XL(V^0\XZ T^\alpha)= -\XL(gV^\alpha{}_\beta T^\beta).
\label{eq:ConstrtR}
\end{equation}
If we can use this relation,  the unwanted 
central-charge transformed quantities 
$\XZ \BF^\alpha_{ab},\,\XZ X^{\alpha ij}$, $\XZ(Z\tau^{\alpha i}),\,\XZ(Z^2\sigma^\alpha)$, 
which are contained in the LHS, 
can immediately be rewritten in terms of the independent variables. 
In order to utilize this relation, we first recall the following facts. 
First, 
${\cal L}_{\rm VL}(V^1\,\XL(V^2\,V^3))$ is trilinear in the three vector 
multiplets $V^1$, $V^2$ and $V^3$, and it is completely symmetric under 
the their interchange {\em if} they are all $G$-neutral (i.e, 
Abelian);
that is, we have the identity
\begin{eqnarray}
{\cal L}_{\rm VL}\left(V^1\,\XL(V^2\,V^3)\right) &=&
{\cal L}_{\rm VL}\left(V^2\,\XL(V^1\,V^3)\right) \nn
&&{}+\bigl[{\cal L}_{\rm VL}\left(V^1\,\XL(V^2\,V^3)\right)
-{\cal L}_{\rm VL}\left(V^2\,\XL(V^1\,V^3)\right)\bigr]_{*\hbox{-}{\rm terms}},
\label{eq:cyclicID}
\end{eqnarray}
where `$*$-terms' indicates all the $G$-transformation terms containing 
the $*$-symbol that are absent 
when the vector multiplets $V$ are Abelian.
When we wish to generalize this identity to cases including tensor 
multiplets, we must {\em define} the quantity 
${\cal L}_{\rm VL}(T\,L)$ for the tensor multiplet $T$, because the VL 
action formula ${\cal L}_{\rm VL}(V\,L)$ explicitly contains the vector 
component $W_\mu$ of $V$, to which the tensor multiplet $T$ has no 
counterpart. However, we recall that the VL action formula can be 
rewritten into a form in which the vector component $W_\mu$ appears 
only in the field strength $F_{\mu\nu}(W)$ {\em if} $V$ and $L$ are 
both $G$-neutral. Thus, 
as a definition of the quantity `${\cal L}_{\rm VL}(T\,L)$', 
we introduce the following function 
${\cal L}_{\rm TL}(T^\alpha\XL(V^IT^\beta))$, which reduces 
to the VL invariant action when the multiplets $T$ and $V^I$ are all 
Abelian vector 
multiplets. Writing $f(\calT)=V^IT^\beta$, we have 
\begin{eqnarray}
&&e^{-1}{\cal L}_{\rm TL}\left(T^\alpha\XL(f(\calT))\right)
\equiv X^{\alpha ij}L_{ij}(f(\calT))
+2i\bar\tau^\alpha\bigl(\varphi(f(\calT))+L(f(\calT))\gamma\dt\psi\bigr) \nn
&&\hspace*{2em}{}+\myfrac12\sigma^\alpha\left(N(f(\calT))
-2i\bar\psi\dt \gamma\varphi(f(\calT))
-2i\bar\psi_a^i\gamma^{ab}\psi_b^jL_{ij}(f(\calT))\right)
+\myfrac14\cBF^\alpha_{\mu\nu}E^{\mu\nu}(f(\calT)),
\nn
&&E^{\mu\nu}(V^IT^\beta)\equiv-\left(
(4v^{\mu\nu}+i\bar\psi_\rho\gamma^{\mu\nu\rho\sigma}\psi_\sigma)M^I\sigma^\beta 
+(F^{\mu\nu I}(W)\sigma^\beta+ M^I\cBF^{\beta\mu\nu}) \right. \nn
&&\qquad \qquad \quad \left.{}
+2i\bar\Omega^I\gamma^{\mu\nu}\tau^\beta 
-2i\bar\psi_\lambda\gamma^{\mu\nu\lambda}(\Omega^I\sigma^\beta+M^I\tau^\beta)
+\myfrac12\epsilon^{\mu\nu\lambda\rho\sigma}W^I_\lambda\cBF^\beta_{\rho\sigma}\right), 
\hspace{1em}
\label{eq:LT}
\end{eqnarray}
where $\cBF_{\mu\nu}^\alpha$ (without a hat) was introduced in
Eq.~(\ref{eq:RQidentity}).  We should, however, note 
a misleading point of our notation in Eq.~(\ref{eq:LT}): 
The function ${\cal L}_{\rm TL}(T^\alpha\XL(V^IT^\beta))$
depends directly on $V^I$ and $T^\beta$ through $E^{\mu\nu},
(V^IT^\beta)$ which is not actually a component of 
the linear multiplet $\XL(V^IT^\beta)$ unless $V^I$ and $T^\beta$ 
are both Abelian vector multiplets. Therefore, for instance, 
although we have the equation $\XL(V^0\XZ T^\alpha)+\XL(gV^\alpha{}_\beta T^\beta)=0$
in Eq.~(\ref{eq:ConstrtR}), we have a non-vanishing difference,
\begin{eqnarray}
&&\hspace*{-3em}{\cal L}_{\rm TL}\left(T^\alpha\XL(V^0\XZ T^\beta)\right)
+{\cal L}_{\rm TL}\left(T^\alpha\XL(gV^\beta{}_\gamma T^\gamma)\right) \nn
&=& \myfrac14\cBF_{\mu\nu}^\alpha 
\left(E^{\mu\nu}(V^0\XZ T^\beta))+E^{\mu\nu}(gV^\beta{}_\gamma T^\gamma)\right) 
= -\myfrac18\epsilon^{\mu\nu\lambda\rho\sigma} \cBF_{\mu\nu}^\alpha\partial_\lambda\cBF_{\rho\sigma}^\beta.
\label{eq:ID2}
\end{eqnarray}
In deriving this, we need to use the previous identity 
(\ref{eq:RQidentity}).
Applying the equations (\ref{eq:ID2}), (\ref{eq:cyclicID}) and 
(\ref{eq:LT}) and the constraint relation (\ref{eq:ConstrtR}), we can 
rewrite the action (\ref{eq:Taction}) as
\begin{eqnarray}
{\cal L}_T&=&-{\cal L}_{\rm VL}\left(V^0 \XL(T^\alpha\XZ T^\beta)\right)d_{\alpha\beta} 
=-{\cal L}_{\rm TL}\left(T^\alpha\XL(V^0\XZ T^\beta)\right)d_{\alpha\beta} 
+ \Delta{\cal L}'(\hbox{$*$-terms}) \nn
&=&
{\cal L}_{\rm TL}\left(T^\alpha\XL(gV^\beta{}_\gamma T^\gamma)\right)d_{\alpha\beta}
\bigr|_{*\hbox{-}{\rm free}}
+\myfrac18\epsilon^{\mu\nu\lambda\rho\sigma} \cBF_{\mu\nu}^\alpha\partial_\lambda\cBF_{\rho\sigma}^\beta d_{\alpha\beta}
+ \Delta{\cal L}(\hbox{$*$-terms}), \nn
&&\hspace*{-2em}\Delta{\cal L}'(\hbox{$*$-terms}) 
\equiv d_{\alpha\beta}\bigl\{
-{\cal L}_{\rm VL}\left(V^0\XL(T^\alpha\XZ T^\beta)\right)\bigr|_{*\hbox{-}{\rm terms}}
+{\cal L}_{\rm TL}\left(T^\alpha\XL(V^0\XZ T^\beta)\right)\bigr|_{*\hbox{-}{\rm terms}}
\bigr\}, \nn 
&&\hspace*{-2em}\Delta{\cal L}(\hbox{$*$-terms}) 
\equiv\Delta{\cal L}'(\hbox{$*$-terms}) 
+{\cal L}_{\rm TL}\left( T^\alpha\XL(gV^\beta{}_\gamma T^\gamma)\right)d_{\alpha\beta}
\bigr|_{*\hbox{-}{\rm terms}} \nn
&&=
d_{\alpha\beta}\bigl\{
-{\cal L}_{\rm VL}\left(V^0\XL(T^\alpha\XZ T^\beta)\right)
\bigr|_{*\hbox{-}{\rm terms}}
+{\cal L}_{\rm TL}\left( T^\alpha\XL(V_*T^\beta)\right)
\bigr|_{**\hbox{-}{\rm terms}}\bigr\},
\end{eqnarray}
where the subs cript `$*$-free' indicates that the $*$-terms are 
all discarded.

Now, this equation is evaluated as follows. Because the 
$*$-free terms are the same as those in the Abelian vector multiplet case, 
the first term is
\begin{equation}
{\cal L}_{\rm TL}\left(T^\alpha\XL(gV^\beta{}_\gamma T^\gamma)\right)d_{\alpha\beta}
\bigr|_{*\hbox{-}{\rm free}}
=
{\cal L}_{\rm VL}\left(gV^\beta{}_\gamma\XL(T^\alpha T^\gamma)\right)d_{\alpha\beta}
\bigr|_{*\hbox{-}{\rm free}}.
\end{equation}
The explicit expression of this term can be directly read from 
our previous result\cite{ref:KO2} for the 
vector multiplet action ${\cal L}_{\rm VL}(V^I\XL(V^JV^K))c_{IJK}$, which 
is solely characterized by the `norm function' 
$\calN(M)=c_{IJK}M^IM^JM^K$. The $\Delta{\cal L}(\hbox{$*$-terms})$ can be directly 
computed by picking out only the terms containing the $*$-symbol in 
${\cal L}_{\rm VL}(V^0\XL(T^\alpha\XZ T^\beta))$ and 
${\cal L}_{\rm TL}(T^\alpha\XL(V_*T^\beta))$ (in fact,
appearing twice in the latter since 
one $*$ is already contained in $V_*T^\beta$).\footnote{%
To avoid possible confusion, we should note that 
these $**$-terms are those in the expression of 
${\cal L}_{\rm TL}(T^\alpha\XL(V_*T^\beta))$
{\em before} the constraint $\XL(V_*T^\beta)=0$ is applied.} 

In this way, we find the following component expression for 
the tensor action (\ref{eq:Taction}):
\begin{eqnarray}
e^{-1}{\cal L}_T&=&
\half\calN\left(
\myfrac12D -\myfrac12i\bar\psi\dt\gamma\chi-\myfrac14R(M)
-\myfrac{i}2\bar\psi_a\gamma^{abc}R_{bc}(Q)+3v^2\right.\nn
&&\qquad \left.{}+i\bar\psi_a\gamma^{abcd}\psi_bv_{cd}-11i\bar\psi_a\psi_bv^{ab}
-6\bar\psi_a\psi_b\bar\psi^a\psi^b+2\bar\psi_a\psi_b\bar\psi_c\gamma^{abcd}\psi_d\right)\nn
&&{}+\half\calN_\etI \left(
i\bar\tau^\etI \chi+i\bar\tau^\etI \gamma\dt R(Q)
+2v^{ab}\cBF_{ab}^\etI 
\right.\nn
&&\qquad \quad \left.{}
-2i\bar\tau^\etI \gamma^{abc}\psi_av_{bc}+6i\bar\tau^\etI \gamma_a\psi_bv^{ab}
\right.\nn
&&\quad \qquad \left.{}
+\myfrac{i}2\bar\psi_a\gamma^{abcd}\psi_b\cBF_{cd}^\etI 
-2i\bar\psi^a\psi^b\cBF_{ab}^\etI 
-4\bar\psi_a\psi_b\bar\psi_c\gamma^{ab}\gamma^c\tau^\etI 
\right)\nn
&&{}-\half\calN_{\etI \etJ }\left(
-\myfrac14 \cBF_{ab}^\etI  \cBF^{ab\etJ }+\myfrac12\pr\calD^a\sigma^\etI 
\pr\calD_a\sigma^\etJ 
+2i\bar\tau^\etI \slash{\calD}'\tau^\etJ +X_{ij}^\etI X^{\etJ ij}\right.\nn
&&\qquad \quad {}-i\bar\tau^\etI \gamma\dt v\tau^\etJ 
+i\bar\psi_a(\gamma\dt \cBF^\etI -2\slash{\calD}' \sigma^\etI )\gamma^a\tau^\etJ \nn
&&\qquad \quad \left.{}
-\bar\psi_a\psi_b\tau^\etI \gamma^{ab}\tau^\etJ 
-2\bar\psi_a\gamma_b\tau^\etI \bar\psi_c\gamma^{ab}\gamma^c\tau^\etJ -2\bar\psi_a\tau^\etI \bar\psi_b\gamma^a\gamma^b\tau^\etJ 
\right)\nn
&&{}+\calN_{\etI \etJ \etK }\left(
i\bar\tau^\etI X^\etJ \tau^\etK 
+\myfrac{i}4\bar\tau^\etI \gamma\dt \cBF^\etJ \tau^\etK 
-\myfrac23\bar\psi_a\gamma_b\tau^\etI \bar\tau^\etJ \gamma^{ab}\tau^\etK 
-\myfrac23\bar\psi^i\dt\gamma\tau^{\etI j}\bar\tau^\etJ _i\tau^\etK _j
\right)\nn
&&{}+\myfrac18e^{-1}\epsilon^{\mu\nu\lambda\rho\sigma}
\cBF_{\mu\nu}^\alpha\calD'_\lambda\cBF_{\rho\sigma}^\beta d_{\alpha\beta} \nn
&&{}+\myfrac14\calN_{\etI \etJ }(gM\sigma)^\etI (gM\sigma)^\etJ  
-2\calN_\etI ig\bar\Omega\tau^\etI 
-\calN_{\etI \etJ }i\bar\tau^\etI gM\tau^\etJ 
+\calN_\etI i\bar\psi\dt \gamma gM\tau^\etI \nn
&&{}+\alpha\left(\alpha^2-(W_0)^2\right)(Z\sigma)^\alpha(Z^2\sigma)^\beta d_{\alpha\beta}
+\myfrac12\left(3\alpha^2-(W_0)^2\right)(Z\sigma)^\alpha(gM Z\sigma)^\beta d_{\alpha\beta}\nn
&&{}+(Z\sigma)^\alpha\left(\alpha W_0^a(\calD_a' Z\sigma)^\beta 
+2i\bar\Omega^0(3\alpha-\slash W^0)(Z\tau)^\beta 
-2\alpha i\bar\psi_a(\alpha+\slash W_0)\gamma^a(Z\tau)^\beta\right)d_{\alpha\beta}\nn
&&{}+2\alpha i(Z\bar\tau)^\alpha(\alpha-\slash W_0)(Z\tau)^\beta d_{\alpha\beta},
\label{eq:TAresult}
\end{eqnarray}
where the `norm function' here is given by
\begin{equation}
\calN \equiv-\sigma^\alpha d_{\alpha\beta}(gM)^\beta{}_\gamma\sigma^\gamma=
-\sigma^\alpha d_{\alpha\beta}\sum_{I\geq1}M^I(gt_I)^\beta{}_\gamma\sigma^\gamma, 
\label{eq:calN0}
\end{equation}
and $\calN_A=\partial\calN(\sigma)/\partial\sigma^A$, 
$\calN_{AB}=\partial^2\calN(\sigma)/\partial\sigma^A\partial\sigma^B$, etc. 
The operator $\calD_a$ is a `homogeneous covariant derivative',
 which is covariant 
under all the homogeneous transformations, $\XM_{ab}$, $\XD$ and $\XU_{ij}$, 
and the gauge transformations $G=G'\times U(1)_Z$, \ and
 $\calD_a'$ is equivalent to $\calD_a$ with $U(1)_Z$ covariantization omitted: 
\begin{eqnarray}
\calD_\mu 
&=& \partial_\mu- \half \omega_\mu^{ab}\XM_{ab} - b_\mu\XD - \half V_\mu^{ij}\XU_{ij}
-W_\mu^0\XZ- \sum_{I\geq1}W_\mu^I\XG_I, \nn
&=& \calD'_\mu-W_\mu^0\XZ.
\end{eqnarray}
The contributions from $\Delta{\cal L}(\hbox{$*$-terms})$ are the terms in the 
last four lines, starting with 
$\myfrac14\calN_{\etI \etJ }(gM\sigma)^\etI (gM\sigma)^\etJ$, aside from 
the $G'$-covariantization terms contained in $\calD'_a$.


Now, let us include the mass terms and consider the general action,
\begin{eqnarray}
{\cal L}_T&=&
-{\cal L}_{\rm VL}\left(V^0 \XL(T^\alpha d_{\alpha\beta}(\XZ T)^\beta)\right)
+{\cal L}_{\rm VL}\left(V^0 \XL(T^\alpha\eta_{\alpha\beta}T^\beta 
+2T^\alpha\eta_{\alpha I}V^I)\right).
\label{eq:FullTaction}
\end{eqnarray} 
Interestingly, the component expression for this case turns 
out to take the same form as the above Eq.~(\ref{eq:TAresult}), provided 
that the following replacements are made. First, 
the central charge transformation is replaced by the new one, 
$\tilde\XZ$:\footnote{Thanks to the 
$G'$-invariance of the tensors $\eta_{\alpha\beta}$ and $\eta_{\alpha I}$ in 
Eq.~(\ref{eq:EtaInv}), this $\tilde\XZ$ transformation 
 commutes with the $G'$ transformation, and therefore it is still
natural to call it the `central transformation'.}
\begin{eqnarray}
\XZ T^\alpha \ &\rightarrow& \ 
\tilde\XZ T^\alpha\equiv \XZ T^\alpha -(gt_0)^\alpha{}_\beta T^\beta- (gt_0)^\alpha{}_I V^I, \nn
&&(gt_0)^\alpha{}_\beta\equiv (d^{-1})^{\alpha\gamma}\eta_{\gamma\beta}, \qquad  
(gt_0)^\alpha{}_I \equiv  (d^{-1})^{\alpha\beta}\eta_{\beta I}.
\label{eq:Ztilde}
\end{eqnarray}
Second, the norm function now reads
\begin{eqnarray}
\calN &\equiv&-\sigma^\alpha d_{\alpha\beta}(gM)^\beta{}_\gamma\sigma^\gamma 
-2\alpha\sigma^\alpha d_{\alpha\beta}(gt_0)^\beta{}_IM^I
=\sigma^\alpha\sigma^\beta M^Ic_{I\alpha\beta}-2\alpha\sigma^\alpha M^I\eta_{\alpha I}\,,\nn
&&c_{I\alpha\beta}\equiv-d_{\alpha\gamma}(gt_I)^\gamma{}_\beta\ .\qquad
(I=0,1,2,\cdots) 
\hspace{3.5em} 
\label{eq:Nresult}
\end{eqnarray}
Note that, in contrast to the previous $\calN$ in (\ref{eq:calN0}), 
 summations over repeated $I$ here and henceforth 
always include $I=0$ 
with $M^0=\alpha$ and $(gt_0)^\alpha{}_A$ defined in Eq.~(\ref{eq:Ztilde}). 
Third, the primed homogeneous covariant derivative $\calD'_\mu$ should 
now be understood to also contain covariantization with respect 
to the `homogeneous part' $\XG_0$ of the central charge transformation 
$\XZ$:
\begin{eqnarray}
\calD'_\mu\equiv \calD_\mu+ W_\mu^0\tilde\XZ
&=& \partial_\mu- \half \omega_\mu^{ab}\XM_{ab} - b_\mu\XD - \half V_\mu^{ij}\XU_{ij}
-W_\mu^0\XG_0- \sum_{I\geq1}W_\mu^I\XG_I, \nn
\XG_0T^\alpha&\equiv& (gt_0)^\alpha{}_\beta T^\beta+(gt_0)^\alpha{}_I V^I.
\label{eq:Dprimerule}
\end{eqnarray}
The group action implied in terms like $gM\sigma^\alpha, \ gM\tau^\alpha$, etc., is 
also understood to contain the `off-diagonal' $(gt_0)^\alpha{}_I$ part:
\begin{equation}
gM\sigma^\alpha= (gM)^\alpha{}_\beta\sigma^\beta+ \alpha(gt_0)^\alpha{}_IM^I.
\end{equation}
Finally, the `Chern-Simons' like term $(e^{-1}/8)\epsilon^{\mu\nu\lambda\rho\sigma}
\cBF_{\mu\nu}^\alpha\calD'_\lambda\cBF_{\rho\sigma}^\beta d_{\alpha\beta}$ is replaced by
\begin{eqnarray}
&&\hspace{-2em}\myfrac18e^{-1}\epsilon^{\mu\nu\lambda\rho\sigma}
\cBF_{\mu\nu}^\alpha\calD''_\lambda\cBF_{\rho\sigma}^\beta d_{\alpha\beta} \nn
&&=\myfrac18e^{-1}\epsilon^{\mu\nu\lambda\rho\sigma}\cBF_{\mu\nu}^\alpha 
\left(
\partial_\lambda B_{\rho\sigma}^\alpha 
-W_\lambda^I(gt_I)^\alpha{}_\beta B_{\rho\sigma}^\beta 
-2W_\lambda^0(gt_0)^\alpha{}_IF_{\rho\sigma}{}^I(W)\right).
\label{eq:CSlike}
\end{eqnarray}
Note the factor 2 multiplying the last covariantization term in 
$\calD''_\lambda\cBF_{\rho\sigma}^\beta$, which differs from the factor 1 in the above 
definitions of $\calD'_\mu$ and the group action $gM$. 

We now explain why these replacement rules appear. First, consider the case 
in which 
there exist no $V$-$T$ mixing mass terms $2T^\alpha\eta_{\alpha I}V^I$. 
Then, the purely tensor mass terms can easily be absorbed into the 
tensor kinetic terms,
\begin{equation}
\calL_T^{\rm pure}=-{\cal L}_{\rm VL}\left(V_0\XL(T^\alpha d_{\alpha\beta}(\XZ T)^\beta-T^\alpha\eta_{\alpha\beta}T^\beta)\right) 
=-{\cal L}_{\rm VL}\left(V_0\XL(T^\alpha d_{\alpha\beta}(\tilde\XZ T)^\beta)\right), 
\label{eq:pureT}
\end{equation}
by using the redefined central charge transformation $\tilde\XZ$ 
given above in Eq.~(\ref{eq:Ztilde}) 
(but with $\eta_{\alpha I}$ set equal to 0 in this case). This is the same form 
as the previous tensor kinetic term, $\calL_T$, in
Eq.~(\ref{eq:Taction}) 
with 
$\XZ$ replaced by $\tilde\XZ$. 
Note, however, that the central charge transformations $\XZ$ contained 
in the $*$-operations of all the other equations, e.g., the constraint 
equation (\ref{eq:Tconstrt}) and the embedding formula (\ref{eq:ZT}), remain
the same as the original $\XZ$. 
We need to rewrite them in terms of the new $\tilde\XZ$. We have
\begin{eqnarray}
&&V_*T^\alpha=V^0(\XZ T)^\alpha+\sum_{I\geq1}V^I(gt_I)^\alpha{}_\beta T^\beta 
=V^0(\tilde\XZ T)^\alpha+V^I(gt_I)^\alpha{}_\beta T^\beta \nn
&&\hbox{with}\quad V^I(gt_I)^\alpha{}_\beta T^\beta=
(V^0(gt_0)^\alpha{}_\beta+\sum_{I\geq1}V^I(gt_I)^\alpha{}_\beta)T^\beta.
\end{eqnarray}
It is thus seen that, after rewriting the central charge, the tensor multiplets $T^\alpha$
become $U(1)_Z$-charged in the usual sense, with generators 
$(t_0)^\alpha{}_\beta$. Therefore, the $G'$-transformation should now include 
$I=0$ everywhere, with $(gt_0)^\alpha{}_\beta=(d^{-1})^{\alpha\gamma}\eta_{\gamma\beta}$, 
and so we obtain the above replacement rules, Eqs.~(\ref{eq:Ztilde})-(\ref{eq:CSlike}), in the case  $(gt_0)^\alpha{}_I=0$. 
Note that the central charge term $\tilde\XZ T$ vanishes
 if equations of motion 
are used. We know that a very similar situation exists also 
for the mass term of the hypermultiplets.\cite{ref:KO2} 

Next, we consider the general action (\ref{eq:FullTaction}), also containing 
the $V$-$T$ transition mass term 
${\cal L}_{\rm VL}\left(V_0\XL(2T^\alpha\eta_{\alpha I}V^I)\right)$. 
This case can in fact be reduced to the pure tensor mass case
considered above as follows. 
Let us consider the pure tensor multiplet system, as in 
Eq.~(\ref{eq:pureT}), in which the tensor multiplets consist of 
three categories: $T^A\equiv(T^I, T^{\bar I}, T^\alpha)$. Here,
$T^I$ is an adjoint representation, $T^{\bar I}$ is another adjoint 
tensor multiplet, and $T^\alpha$ is the rest, which can be of arbitrary 
representation. We chose invariant $d$ and $\eta$ tensors in the forms
\begin{equation}
d_{AB}= \pmatrix{
0 & d_{I\bar J} & 0 \cr
d_{\bar IJ}=-d_{J\bar I} & 0 & 0\cr
0 & 0 & d_{\alpha\beta}\cr} = -d_{BA}, \qquad 
\eta_{AB}= \pmatrix{
0 & 0 & \eta_{I\beta} \cr
0 & 0 & 0 \cr
\eta_{\alpha J} & 0 & \eta_{\alpha\beta}\cr}
=\eta_{BA}.
\end{equation}
Then the action (\ref{eq:pureT}) becomes
\begin{eqnarray}
{\cal L}'_T
&=&-{\cal L}_{\rm VL}\left(V_0\XL(T^\alpha d_{\alpha\beta}(\XZ T)^\beta 
+T^Id_{I\bar J}(\XZ T)^{\bar J}
+T^{\bar I}d_{\bar I J}(\XZ T)^J)\right) \nn
&&\hspace{2em}{}+{\cal L}_{\rm VL}\left(V_0\XL(
T^\alpha\eta_{\alpha\beta}T^\beta+2T^\alpha\eta_{\alpha I}T^I)\right). 
\label{eq:Taction2}
\end{eqnarray}
Note that we have included no mass terms containing $T^{\bar I}$.
Then, this system is a pure tensor multiplet system in any case, and therefore 
the previous result for the component expression 
(\ref{eq:TAresult}) of the action 
applies with replacement rules given in Eqs.~(\ref{eq:Ztilde})-(\ref{eq:CSlike}), with 
$(\eta_{\alpha\beta},\eta_{\alpha I}) \rightarrow (\eta_{AB}, 0)$
understood in this case. 
Now, we stipulate that the first adjoint tensor multiplet $T^I$ be  
a vector multiplet $V^I$; i.e., We set $T^I=V^I$. 
This is allowed, because the vector multiplet is a 
special tensor multiplet (which is $\XZ$-invariant), and all the 
manipulations in computing the above component expression remain 
valid also for vector multiplets. 
Then, the action (\ref{eq:Taction2}) of this system clearly reduces to 
the desired general action (\ref{eq:FullTaction}), because 
$\XZ V^I =0$ and $
{\cal L}_{\rm VL}\left(V_0\XL(V^Id_{I\bar J}(\XZ T)^{\bar J})\right)=
{\cal L}_{\rm VL}\left(V_0\XZ\XL(V^Id_{I\bar J}T^{\bar J})\right)$ 
vanishes up to total derivatives. This also implies that the action 
becomes completely independent of the adjoint tensor multiplet 
$T^{\bar I}$. 

We now apply the above replacement rules. The central charge 
transformation is replaced by $\tilde\XZ T^A=\XZ T^A - (gt_0)^A{}_BT^B$. 
Because we have  
\begin{equation}
(gt_0)^A{}_B
\equiv (d^{-1})^{AC}\eta_{CB}= 
\pmatrix{
0 & 0 & 0 \cr
0 & 0 & (d^{-1})^{\bar IK}\eta_{K\beta} \cr
(d^{-1})^{\alpha\gamma}\eta_{\gamma J} & 0 & (d^{-1})^{\alpha\gamma}\eta_{\gamma\beta}\cr},
\quad 
T^A=\pmatrix{ V^I \cr T^{\bar I} \cr T^\alpha\cr},
\label{eq:gt0}
\end{equation}
 $\tilde\XZ T^\alpha$ actually reproduces the rule (\ref{eq:Ztilde}),
and also
\begin{equation}
\tilde\XZ V^I=\XZ V^I  = 0, \qquad 
\tilde\XZ T^{\bar I}=\XZ T^{\bar I} - (d^{-1})^{\bar IK}\eta_{K\beta}T^\beta.
\end{equation}
Accordingly, the primed derivative $\calD'_\mu$ applied to $T^\alpha$ also reproduces the 
rule (\ref{eq:Dprimerule}), and $\calD'_\mu$ applied to $V^I$ 
undergoes no change:
$\calD'_\mu V^I=\calD_\mu V^I$. 
Applying the derivative $\calD'_\mu$ to $T^{\bar I}$ yields
\begin{equation}
\calD'_\mu T^{\bar I}=
\calD^{\prime {\rm hom}}_\mu T^{\bar I}
-W^0_\mu(d^{-1})^{\bar IK}\eta_{K\beta}T^\beta, 
\label{eq:Dprime2}
\end{equation}
where $\calD^{\prime {\rm hom}}_\mu T^{\bar I}$ denotes the part 
homogeneous in $T^{\bar I}$. 
Although it is guaranteed that the component fields of $T^{\bar I}$ 
never appear in the action, the second term 
$-W^0_\mu(d^{-1})^{\bar IK}\eta_{K\beta}T^\beta$ may give a nonvanishing 
contribution. This in fact happens only 
in Chern-Simons-like terms, which now read
\begin{eqnarray}
&&\hspace{-1em}\myfrac18e^{-1}\epsilon^{\mu\nu\lambda\rho\sigma}
\cBF_{\mu\nu}^A\calD'_\lambda\cBF_{\rho\sigma}^Bd_{AB} \nn
&&{}=
\myfrac18e^{-1}\epsilon^{\mu\nu\lambda\rho\sigma}
\bigl\{
\cBF_{\mu\nu}^\alpha\calD'_\lambda\cBF_{\rho\sigma}^\beta d_{\alpha\beta} 
+F_{\mu\nu}^I(W)\calD'_\lambda\cBF_{\rho\sigma}^{\bar J}d_{I{\bar J}} 
+\cBF_{\mu\nu}^{\bar J}\calD'_\lambda F_{\rho\sigma}^I(W)d_{{\bar J}I} 
\bigr\}
\label{eq:532}
\end{eqnarray}
Note that in the second term, $\calD'_\lambda\cBF_{\rho\sigma}^{\bar J}=
\calD^{\prime {\rm hom}}_\lambda\cBF_{\rho\sigma}^{\bar J}
-W^0_\lambda(d^{-1})^{{\bar J}K}\eta_{K\alpha} \cBF_{\rho\sigma}^\alpha$ 
by Eq.~(\ref{eq:Dprime2}).
The $\calD^{\prime {\rm hom}}_\lambda\cBF_{\rho\sigma}^{\bar J}$ part, 
after performing the partial integration, gives the same 
contribution as the third term, which vanishes by the Bianchi identity 
$\epsilon^{\mu\nu\lambda\rho\sigma}\calD_\lambda F_{\rho\sigma}^I(W)d_{{\bar J}I}=0$. 
The contribution of the remaining part, 
$-W^0_\lambda(d^{-1})^{{\bar J}K}\eta_{K\alpha} \cBF_{\rho\sigma}^\alpha$, in 
$\calD'_\lambda\cBF_{\rho\sigma}^{\bar J}$ is seen to give 
\begin{equation}
-\myfrac18e^{-1}\epsilon^{\mu\nu\lambda\rho\sigma}\left\{
F_{\mu\nu}^I(W)W^0_\lambda\eta_{I\alpha} \cBF_{\rho\sigma}^\alpha 
=
\cBF_{\mu\nu}^\alpha W^0_\lambda\eta_{\alpha I} F_{\rho\sigma}^I(W)
=
\cBF_{\mu\nu}^\alpha W^0_\lambda d_{\alpha\beta}(gt_0)^\beta{}_I F_{\rho\sigma}^I(W)
\right\}.
\end{equation}
This doubles the 
last covariantization term, $\cBF_{\mu\nu}^\alpha 
\left(-W_\lambda^0(gt_0)^\alpha{}_IF_{\rho\sigma}{}^I(W)\right)
d_{\alpha\beta}$, of the first term contribution in Eq.~(\ref{eq:532}),
\begin{equation}
\myfrac18e^{-1}\epsilon^{\mu\nu\lambda\rho\sigma}
\cBF_{\mu\nu}^\alpha 
\bigl\{\calD'_\lambda\cBF_{\rho\sigma}^\beta 
=\left(
\partial_\lambda B_{\rho\sigma}^\alpha 
-W_\lambda^I(gt_I)^\alpha{}_\beta B_{\rho\sigma}^\beta 
-W_\lambda^0(gt_0)^\alpha{}_IF_{\rho\sigma}{}^I(W)\right)
\bigr\} d_{\alpha\beta},
\end{equation}
and actually reproduces the replacement rule in Eq.~(\ref{eq:CSlike}).

Finally, the norm function now is given by
\begin{eqnarray}
\calN \equiv-\sigma^Ad_{AB}(gM)^B{}_C\sigma^C 
&=&
-\sigma^\alpha d_{\alpha\beta}\bigl((gM)^\beta{}_\gamma\sigma^\gamma 
+\alpha(gt_0)^\beta{}_IM^I\bigr) \nn
&&\hspace{-6em}{}-M^Id_{I\bar K}\bigl((gM)^{\bar K}{}_{\bar J}\sigma^{\bar J}
+\alpha(gt_0)^{\bar J}{}_\alpha\sigma^\alpha\bigr)
-\sigma^{\bar J}d_{{\bar J}K}(gM)^K{}_IM^I.
\end{eqnarray}
However, because 
$M^Id_{I\bar K}(gM)^{\bar K}{}_{\bar J}\sigma^{\bar J}
=\sigma^{\bar J}d_{{\bar J}K}(gM)^K{}_IM^I$, due to the $G'$-invariance of 
$M^Id_{I\bar J}\sigma^{\bar J}$ and the relation
$(gM)^K{}_IM^I= [gM,\,M]^K=0$, this reduces to
\begin{eqnarray}
\calN =
-\sigma^\alpha d_{\alpha\beta}(gM)^\beta{}_\gamma\sigma^\gamma 
-\alpha\sigma^\alpha d_{\alpha\beta}(gt_0)^\beta{}_IM^I
-\alpha M^Id_{I\bar K}(gt_0)^{\bar J}{}_\alpha\sigma^\alpha.\label{eq:calNm}
\end{eqnarray}
If we further substitute the expressions for $(gt_0)^\beta{}_I$ and 
$(gt_0)^{\bar J}{}_\alpha$ given in Eq.~(\ref{eq:gt0}) 
into Eq.~(\ref{eq:calNm}), the 
last two terms on the RHS give $2\alpha\sigma^\alpha\eta_{\alpha
I}M^I$,
 and so this $\calN$ 
reproduces that given in (\ref{eq:Nresult}). This completes the 
proof of the replacement rules given in  
Eqs.~(\ref{eq:Ztilde}) - (\ref{eq:CSlike})

A few comments are in order concerning the result (\ref{eq:TAresult}). 

i) First, the action of the tensor multiplets 
 actually gives a contribution to the scalar potential
$V_{\rm scalar}$: 
\begin{equation}
V_{\rm scalar}=- \myfrac14\calN_{\etI \etJ }(gM\sigma)^\etI
(gM\sigma)^\etJ, 
\end{equation}
Note that $gM\sigma^A$ vanishes for $A=I$, because $gM\sigma^I=[gM,M]^I=0$. 
Since $gM\sigma^\alpha= M^I(gt_I)^\alpha{}_\beta\sigma^\beta+ \alpha(gt_0)^\alpha{}_JM^J$,
this potential contains terms quadratic, cubic and quartic in 
$M^I\ (I=0,1,2,\cdots)$ (recall that $M^0=\alpha$).
Only the first terms quadratic in the $M$ (which are also quadratic in
the $\sigma$) 
was found by G\"unaydin and Zagermann,\cite{ref:GZ} while the other terms 
exist only when the tensor-vector mixing $(gt_0)^\alpha{}_J$ is introduced. 
Note, however, that this mixing comes from the tensor-vector mixing 
{\em mass term}. 
The general tensor-vector mixing terms 
$(gt_I)^\alpha{}_J$ for $I\not=0$ in the 
$G'$ transformation, as was introduced by Bergshoeff 
et~al.,\cite{ref:BCWGHVP} can be eliminated by the field redefinitions, 
as we saw above, and then play no role here.

ii) The action (\ref{eq:TAresult}) does not contain the kinetic terms 
for the vector multiplets $V^I$, because we have inserted the central 
charge transformation operator $\XZ$ in the initial Lagrangian 
${\cal L}_{\rm VL}(V^0L(T^\alpha\XZ T^\beta)d_{\alpha\beta})$. As done in 
Ref.~\citen{ref:KO2}, the general kinetic terms for the vector multiplets 
can be given by an action of the 
form\footnote{%
Precisely speaking, the VL action formula ${\cal L}_{\rm VL}(V^IL(V^JV^K))$ 
applies only to the case in which the index $I$ is that of an Abelian vector 
multiplet. Nevertheless, one can construct an invariant action 
corresponding to the form (\ref{eq:Vaction}) as long as 
$c_{IJK}V^IV^JV^K$ is $G'$-invariant. (See Ref.~\citen{ref:KO2} for 
details.)}
\begin{equation}
-{\cal L}_{\rm VL}(V^IL(V^JV^K))c_{IJK}.
\label{eq:Vaction}
\end{equation}
If we add this action of vector kinetic terms to the above tensor action
(\ref{eq:Taction}), the resultant component expression is still 
given by (\ref{eq:TAresult}), provided that 
the `norm function' $\calN$ there is understood to be
\begin{equation}
\calN \equiv c_{I\alpha\beta}M^I\sigma^\alpha\sigma^\beta+2\eta_{\alpha I}\alpha M^I\sigma^\alpha 
+ c_{IJK}M^IM^JM^K,
\end{equation}
and the following Chern-Simons term is added to the previous Chern-Simons-like 
term (\ref{eq:CSlike}):
\begin{eqnarray}
{\cal L}_{\rm C\hbox{-}S}&=&
\frac{1}8 c_{IJK}\epsilon^{\lambda\mu\nu\rho\sigma}W_{\lambda}^I
\bigl(F_{\mu\nu}^J(W)F_{\rho\sigma}^K(W) 
+\myfrac{1}2g[W_\mu,W_\nu]^JF_{\rho\sigma}^K(W) \nn
&&\hspace{8em}{}+\myfrac{1}{10}g^2[W_\mu,W_\nu]^J[W_\rho,W_\sigma]^K\bigr)\,. 
\label{eq:ChernSimons}
\end{eqnarray}

iii) As seen from the action (\ref{eq:TAresult}), the tensor fields 
$B_{\mu\nu}^\alpha$ posses the first-order kinetic term (\ref{eq:CSlike}), as 
well as the mass term, and therefore they are `self-dual' tensor fields 
of the type (\ref{eq:selfdual}), as discussed in Ref.\citen{ref:PTvN}. 
The number of the on-shell modes of $B_{\mu\nu}^\alpha$ for each 
$\alpha$ is ${}_{(5-1)}C_2/2=3$. We also see that $Z\sigma,\ Z^2\sigma$ 
and $Z\tau^i$ are all non-propagating auxiliary fields, so that the 
on-shell modes in the large tensor multiplet $T^\alpha$ for each 
$\alpha$ are $1(\sigma)+3(B_{\mu\nu})$ bosons plus $4(\tau^i)$
fermions.
The on-shell and off-shell mode counting is summarized in Table
\ref{table:2} for the three multiplets: large tensor multiplet, tensor 
gauge multiplet and vector multiplet.  
Note also that the number of these large tensor multiplets $T^\alpha$ 
appearing in the action is always {\em even}, because the coefficient 
$d_{\alpha\beta}$ of the kinetic term must be antisymmetric under 
$\alpha\leftrightarrow \beta$. (The symmetric part of $d_{\alpha\beta}$, 
if any, could add only a total derivative term to the Lagrangian.)

\begin{table}[tb]
\caption{Comparison of the multiplets. 
The numbers outside and inside the parentheses denote the off-shell
 and on-shell degrees of freedom, respectively, of each component field.  }
\label{table:2}
\begin{center}
\begin{tabular}{l|cc|cc|cc} \hline \hline
multiplets
 &large tensor  ${\mbf T}$& 16+16   & tensor gauge ${\mbf A}$& 8+8 
&vector ${\mbf V}$& 8+8       \\ \hline
 constraints 
&\multicolumn{2}{c|}{${\mbf L(V_*T)}=0$}    
&\multicolumn{2}{c|}{${\mbf L}({\mbf V}^0{\mbf A})=0$} 
&\multicolumn{2}{c}{${\mbf ZV}=0$}        \\ \hline
&$\sigma$     & 1(1)  &$\sigma$      &1(1)  &$M$         &1(1)  \\ \cline{2-7} 
&$\tau^i$    & 8(4)  &$\tau^i$     &8(4)  &$\Omega^i$ &8(4) \\ \cline{2-7}
&$\hat B_{ab}$& 10(3) &$A_{\mu\nu}$&6(3)  &$W_\mu$     &4(3)  \\  
components
&$X^{ij}$     & 3(0)  &              &      &$Y^{ij}$    &3(0)\\ 
&$Z\sigma$    & 1(0)  &$Z\sigma$     &1(0)  &            &    \\ \cline{2-7}
&$Z\tau^i$   & 8(0)  &              &      &            &    \\ \cline{2-7}
&$Z^2\sigma$  & 1(0)  &              &      &            &      \\ \hline
\end{tabular}
\end{center}
\end{table}

\section{An invariant action for $\GT$ and a duality relation}

The tensor gauge (small tensor) multiplet is just a special tensor multiplet 
that satisfies the stronger constraint (\ref{eq:TGconstrt}), so that 
we can apply to it the embedding into the linear multiplet formula 
(\ref{eq:ZT}) and then the VL action formula, to obtain an invariant action 
for a tensor gauge multiplet $\GT$:
\begin{eqnarray}
{\cal L}_\GT&=&-{\cal L}_{\rm VL}(V^0 \XL(\GT \,\GT )).
\label{eq:Baction}
\end{eqnarray}
When computing the explicit form of this action, we need to use 
the complicated component expression of the constraints 
(\ref{eq:TGconstrt}). However, we can avoid this,as in the 
large tensor multiplet case of the previous section. Since,
 for any quantity $X$, 
\begin{eqnarray}
X = X\bigr|_{*\hbox{-}{\rm free}}
+X\bigr|_{*\hbox{-}{\rm terms}}
\end{eqnarray}
holds trivially, we apply it to 
$X=-{\cal L}_{\rm VL}(V^0\XL(\GT \,\GT ))+2{\cal L}_{\rm TL}(\GT \XL(V^0\GT ))$ and 
use the identity
\begin{equation}
{\cal L}_{\rm VL}(V^0\XL(\GT \,\GT ))
\bigr|_{*\hbox{-}{\rm free}}
={\cal L}_{\rm TL}(\GT \XL(V^0\GT ))\bigr|_{*\hbox{-}{\rm free}},
\end{equation}
as before. Then we have
\begin{eqnarray}
&&{\cal L}_\GT  =
{\cal L}_{\rm VL}(V^0\XL(\GT \,\GT ))\bigr|_{*\hbox{-}{\rm free}}
-2{\cal L}_{\rm TL}(\GT \XL(V^0\GT ))
+\Delta{\cal L}(\hbox{$*$-terms}),  \nn
&&\Delta{\cal L}(\hbox{$*$-terms}) \equiv 
[2{\cal L}_{\rm TL}(\GT \XL(V^0\GT ))-{\cal L}_{\rm VL}(V^0\XL(\GT \,\GT ))]
\bigr|_{*\hbox{-}{\rm terms}}.
\label{eq:Baction2}
\end{eqnarray}
The reason we have considered the combination
$2{\cal L}_{\rm TL}(\GT \XL(V^0\GT ))-{\cal L}_{\rm VL}(V^0\XL(\GT \,\GT ))$ is that 
the $\XZ(Z\sigma)$ and $\XZ\tau$ terms are contained in
${\cal L}_{\rm VL}(V^0\XL(\GT \,\GT ))$ twice as many times 
as in ${\cal L}_{\rm TL}(\GT \XL(V^0\GT ))$.

To see the content of this action, let us examine only the 
bosonic part for simplicity. With the aid of Eq.~(\ref{eq:Baction2}), 
the bosonic part can be read as follows:
\begin{eqnarray}
e^{-1}{\cal L}_\GT \bigr|_{\rm boson}&=&
\alpha\left(\myfrac14\cBF_{ab}\cBF^{ab}+\myfrac12\calD'_a\sigma\calD'^a\sigma 
-\left(\frac{\sigma}{\alpha}\right)^2Y_{0ij}Y_0^{ij}\right)
+\myfrac12\calD'_a(\sigma^2)\,\calD^a\alpha 
\nn
&&{}+(\myfrac18R(M)-\myfrac14D-\myfrac32v^2)\alpha\sigma^2
-F_{ab}(W^0)v^{ab}\,\sigma^2 \nn
&&{}+\myfrac12\alpha(\alpha^2-(W_b^0)^2)(Z\sigma)^2
+\myfrac18e^{-1}\epsilon^{\lambda\mu\nu\rho\sigma}W_\lambda^0\cBF_{\mu\nu}\cBF_{\rho\sigma}.
\end{eqnarray}
This expression is not yet the final one, since  $\cBF_{\mu\nu}$ 
should be rewritten in terms of the tensor gauge field $\AB_{\mu\nu}$. 
As derived in the Appendix, the terms containing $\cBF_{\mu\nu}$ are written as
\begin{eqnarray}
&&
\myfrac14\alpha\BF_{ab}\BF^{ab}+\myfrac18\epsilon^{abcde}W_a^0\BF_{bc}\BF_{de}
\nn 
&&=\frac1{4\alpha}\left\{ {1\over3}(\FA _{abc})^2 
+{1\over\alpha^2-(W^0)^2}\bigl[\left(\FA _{abc}W^{0c}\right)^2
-\myfrac16\epsilon^{abcde}\alpha\FA _{abc}\FA _{def}W^{0f}
\bigr]\right\}\hspace{2em}
\label{eq:solsol}
\end{eqnarray}
where $\FA _{abc}$ is the quantity introduced in Eq.~(\ref{eq:delta}), and 
the boson part is essentially the field strength of the tensor 
gauge field $\AB_{\mu\nu}$:
\begin{equation}
\FA _{\lambda\mu\nu}\bigr|_{\rm boson}\equiv 
3\partial_{[\lambda}\AB_{\mu\nu]}
-\myfrac12\epsilon_{\lambda\mu\nu\rho\sigma}
\left(4v^{\rho\sigma}\alpha\sigma+\hat F^{\rho\sigma}(W^0)\sigma\right).
\end{equation}
Thus we see that ${\cal L}_\GT $ contains the kinetic terms of the tensor gauge 
field $\AB_{\mu\nu}$ as well as the scalar field $\sigma$ correctly. 

Finally, we show that this tensor gauge multiplet is dual to 
the vector multiplet. To show this, let us consider the 
following action for a tensor gauge multiplet $\GT $ and a large tensor 
multiplet $T$ which is $G'$-neutral:
\begin{equation}
{\cal L}={\cal L}_{\rm VL}(V^0\XL(T^2))+2{\cal L}_{\rm VL}(V^0\XL(\GT T)),
\label{eq:Daction}
\end{equation}
If we first integrate out $\GT $, or equivalently use equations of motion 
resulting from $\delta S/\delta\GT =0$, we find that the tensor multiplet $T$ 
reduces to a vector multiplet $V$. 
This can be seen as follows. We attach the superscript $T$ to the component 
fields of the large tensor multiplet as 
$T=(\sigma^T, \tau^T, \cBF_{ab}^T, Z\sigma^T,\cdots)$ 
to distinguish them from those of the tensor gauge multiplet 
$\GT=(\sigma, \tau, \AB_{\mu\nu}, Z\sigma)$. 
If we concentrate on the bosonic parts, we have
\begin{eqnarray}
{\cal L}_{\rm VL}(V^0\XL(\GT  T))&\sim&
-\myfrac18\epsilon^{\mu\nu\lambda\rho\sigma}\cBF_{\mu\nu}^T\partial_\lambda\AB_{\rho\sigma}
-\myfrac12\alpha(\alpha^2-(W^0)^2)(Z\sigma^T)
(Z\sigma) \nn
&&{}+\myfrac12\alpha\left[(\alpha^2-(W^0)^2)(Z^2\sigma^T)
+(\hbox{terms containing }Z\sigma^T)\right]\sigma, 
\hspace{3em}
\end{eqnarray}
so that the equations of motion $\delta S/\delta\GT =0$ yield
\begin{equation}
\left({\delta S\over\delta(Z\sigma)},\ {\delta S\over\delta\sigma},
\ {\delta S\over\delta\AB_{\mu\nu}}\right)=0\ \Rightarrow\ 
\left(Z\sigma^T,\ Z^2\sigma^T, 
\ \epsilon^{\mu\nu\lambda\rho\sigma}\partial_\lambda\cBF_{\mu\nu}^T\right)=0.
\end{equation}
Thus, the tensor multiplet has vanishing central charge, and $\cBF_{\mu\nu}$ 
satisfies the Bianchi identity, so that it can be expressed by a vector 
field. The fermionic part should also reduce to that of a vector multiplet 
by supersymmetry.  If we substitute the solution $T=V$ back into the 
action (\ref{eq:Daction}), then, using the identity 
${\cal L}_{\rm VL}(V^0\XL(\GT V))=0$, we obtain 
\begin{equation}
{\cal L}={\cal L}_{\rm VL}(V^0\XL(VV)),
\end{equation}
which is the action for the vector multiplet 
$V$. [All the component fields of the central charge vector multiplet $V^0$, 
except the gravi-photon, are eliminated by the dilatation $\XD$ and special 
supersymmetry $\XS$ gauge fixing or, otherwise, by non-propagating auxiliary
fields.]\  

It may be necessary to add an explanation of the identity ${\cal L}_{\rm 
VL}(V^0\XL(\GT V))=0$. This follows from the more general identity 
\begin{eqnarray}
{\cal L}_{\rm VL}(V^0\XL(V_1T))={\cal L}_{\rm VL}(V_1\XL(V^0T)),
\label{eq:id3}
\end{eqnarray}
which holds for any $G'$-neutral (Abelian) vector multiplet $V_1$ and 
any tensor multiplet, a large one $T$ or a small one $\GT $. 
Then, applying this, we have
\begin{eqnarray}
{\cal L}_{\rm VL}(V^0\XL(\GT V))={\cal L}_{\rm VL}(V\XL(V^0\GT ))=0,
\end{eqnarray}
using the constraint $\XL(V^0\GT )=0$.

If we instead integrate $T$ out first in the initial action 
(\ref{eq:Daction}), then we obtain 
\begin{equation}
{\cal L}={\cal L}(V^0\XL((T+\GT )^2))-{\cal L}(V^0\XL(\GT \GT )) \Rightarrow-{\cal L}(V^0\XL(\GT \GT )),
\end{equation}
the action for a tensor gauge multiplet. The first term, 
${\cal L}(V^0\XL((T+\GT )^2))$,represents the non-propagating `mass' terms for 
the tensor multiplet $T'\equiv T+\GT $ and can be eliminated.

This duality can be shown in the opposite way if we start from the action
\begin{equation}
{\cal L}=-{\cal L}(V^0\XL(T^2))+2{\cal L}(V^0\XL(VT)).
\label{eq:Daction2}
\end{equation}
Then, integrating $V$ out first yields the equation of motion 
$\XL(V^0T)=0$ by (\ref{eq:id3}), so that the tensor multiplet $T$ reduces
to a tensor gauge multiplet $\GT $. Substituting this solution 
$T=\GT$ back into the action yields the tensor gauge multiplet action 
$-{\cal L}(V^0\XL(\GT \GT ))$, since ${\cal L}(V^0\XL(V\GT ))=0$. 
Integrating out $T$ 
first, on the other hand, gives the vector action ${\cal L}(V^0\XL(VV))$.

%

\section*{Acknowledgements}
The authors would like to thank Sorin Cucu, Stefan Vandoren and 
Antoine Van Proeyen for valuable discussions.
T.~K.\ is supported in part by a Grant-in-Aid for Scientific Research 
(No.~13640279) from the Japan Society for the Promotion of Science.
T.~K.\ and K.~O.\ are supported in part by a 
Grant-in-Aid for Scientific Research on Priority Areas (No.~12047214)
and a Grant-in-Aid for Scientific Research (No.~01350), respectively, 
from the Ministry of Education, Culture, Sports, Science and Technology, Japan.
The work of  K.~O.\ is supported in part by  the Japan Society for the
Promotion of Science under the Pre-doctoral Research Program.

\appendix

\section{Solving Eq.~(\protect\ref{eq:bianchi})}

Let us solve an equation of the form (\ref{eq:bianchi}): 
\begin{equation}
3W_{[a}\BF_{bc]}+\myfrac12M\epsilon_{abcde}\BF^{de} =
\FA _{abc}.
\label{eq:A1}
\end{equation}

We define the following two operations 
$\calP_\|$ and $\calR_\perp$ on rank 2 antisymmetric tensors $T_{ab}$:
\begin{eqnarray}
\calP_\|T_{ab} &\equiv& 
\myfrac{W_aW^c}{W^2}T_{cb}+T_{ac}\myfrac{W^cW_b}{W^2},\nn
\calR_\perp T_{ab} &\equiv& \myfrac1{2M}\epsilon_{abcde}W^cT^{de}.
\end{eqnarray}
Then $\calP_\|$ and 
$(M^2/W^aW_a)\calR_\perp^2$ are 
`longitudinal' and transverse projection operators, respectively:
\begin{eqnarray}
&&\calP_\|^2=\calP_\|,\qquad 
\lambda^{-1}\calR_\perp^2 + \calP_\| = 1,\qquad 
\calR_\perp\calP_\|=0,\qquad \lambda\equiv\myfrac{W^2}{M^2}.
\end{eqnarray}
Taking the dual of the given equation (\ref{eq:A1}), 
we rewrite it as
\begin{eqnarray}
\tilde \FA _{ab}&=&M\BF_{ab}+\myfrac12\epsilon_{abcde}W^c\BF^{de}
=M(1+\calR_\perp)\BF_{ab}.
\end{eqnarray}
Then we can solve it as
\begin{eqnarray}
\BF_{ab}&=&\myfrac1M(1+\calR_\perp)^{-1}\tilde \FA _{ab}
=\myfrac1{M(1-\lambda)}(1-\calR_\perp-\lambda\calP_\|)\tilde \FA _{ab},
\end{eqnarray}
yielding Eq.~(\ref{eq:sol}). Using this we have
\begin{eqnarray}
&&\hspace*{-3em}
\myfrac14M\BF_{ab}\BF^{ab}+\myfrac18\epsilon^{abcde}W_a\BF_{bc}\BF_{de}\nn
&=&\myfrac14M\BF_{ab}(1+\calR_\perp)\BF^{ab}
=\myfrac1{4M}\tilde \FA _{ab}(1+\calR_\perp)^{-1}\tilde \FA ^{ab}\nn
&=&\myfrac1{4(M^2-W^2)}\tilde \FA _{ab}(1-\calR_\perp-\lambda\calP_\|)\tilde \FA ^{ab},
\end{eqnarray}
which gives Eq.~(\ref{eq:solsol}) in the text.

\end{document}